\documentclass{article}

\newtheorem{definition}{Definition}

\usepackage{amssymb}
\usepackage{amsmath}
\usepackage{graphicx}
\usepackage[tight,footnotesize]{subfigure}
\usepackage{algorithm}
\usepackage{algorithmic}

\begin{document}

\title{A generic trust framework for large-scale open systems using machine learning}


\author{Xin Liu\\Nanyang Technological University\\liu\_xin@pmail.ntu.edu.sg \and
Gilles Tredan\\Technische Universit\"at Berlin/T-labs\\gilles@net.t-labs.tu-berlin.de \and
Anwitaman Datta\\Nanyang Technological University\\anwitaman@ntu.edu.sg \and}

\date{}
\maketitle

%
%




\begin{abstract}

In many large scale distributed systems and on the web, agents need to interact with other unknown agents to carry out some tasks or transactions. The ability to reason about and assess the potential risks in carrying out such transactions is essential for providing a safe and reliable environment.
A traditional approach to reason about the trustworthiness of a transaction is to determine the trustworthiness of the specific agent involved, derived from the history of its behavior.
As a departure from such traditional trust models, we propose a generic, machine learning approach based trust framework where an agent uses its own previous transactions (with other agents) to build a knowledge base, and utilize this to assess the trustworthiness of a transaction based on associated features, which are capable of distinguishing successful transactions from unsuccessful ones. These features are harnessed using appropriate machine learning algorithms to extract relationships between the potential transaction and previous transactions. The trace driven experiments using real auction dataset show that this approach provides good accuracy and is highly efficient compared to other trust mechanisms, especially when historical information of the specific agent is rare, incomplete or inaccurate.\\
\textbf{Keywords:} trust management, machine learning, features, large-scale systems
\end{abstract}

%
%



\section{Introduction}
\label{sec:introduction}

Trust is an important ingredient facilitating reliable
interactions\footnote{Throughout this paper, we use the terms interaction and transaction interchangeably.} among autonomous participants without global
coordination in diverse large-scale systems including e-commence, distributed
and peer-to-peer systems, multi-agent systems and dynamic collaborative systems. Due to
the large scale and openness of these systems, an agent is often required to
interact with other agents with which it has few or no shared past
interactions. To assess the risk of such interactions and to determine whether
an unknown agent is trustworthy, an efficient trust mechanism is necessary
\cite{josongsurvey, canwetrusttrust}.

Traditional approaches, while effective when the necessary information is available,
often rely upon knowledge that may not actually be available locally to the assessor. For instance, the interaction partner recently joins the system so its past behavior is quite limited; or it is difficult to form the trust path to derive trustworthiness of the target agent (i.e., opinions from other agents who have interacted with the target agent are insufficient or ingenuine). Hence, we want to develop a new trust mechanism to compute trust even in absence of historical information of the specific agent in question.
In this paper, we propose a generic trust framework, which investigates trustor's local knowledge using machine learning algorithms. Machine learning aims to automatically learn to recognize complex patterns and make intelligent decisions based on existing datasets. Since each agent may have past transactions with other agents\footnote{One weak assumption of this work is that trustor has sufficient past transactions such that machine learning algorithms can be performed. In case trustor is new or inactive in the system, i.e., no or few past transactions are available, we use a overlay network where peers share their local knowledge to augment the trustor's sparse knowledge.}, we argue that by investigating useful features that are capable of distinguishing successful transactions from unsuccessful ones, we can apply sophisticated machine learning algorithms to analyze past transactions to learn what a successful transaction is and then apply the result of learning to predict trustworthiness of a potential transaction.
This new mechanism, given its use of different kind of information, is meant to complement traditional models, and given its reliance of different set of information, should and can not be compared beyond the purpose of validation.


To explain how our approach works and differs from
traditional trust mechanisms, let us use an analogy with our behavior in the
real world. When a customer, say Bob, wants to evaluate the trustworthiness of
a provider, say Sally's restaurant (i.e., to guess how he will be satisfied by Sally's
food), it is possible to consider two different cases:
\begin{itemize}
\item Bob has access to some feedbacks specifically about Sally's restaurant. For instance,
  Bob has already eaten there in the past, or maybe one of Bob's friends did,
  and gave Bob some feedbacks about it. Then Bob uses these feedbacks to decide
  whether he will eat at Sally's restaurant or not.
  This is the case that traditional recommendation systems rely upon. More
  generally these recommendation systems will try to find feedbacks about
  Sally's restaurant among Bob's friends, or among the friends of Bob's friends
  (FoF), or from online restaurant review sites and so on. As we will see in the related works section, this method
  performs well, but has several drawbacks: \emph{(i)} Bob has to find somehow a
  chain of friends that will provide him feedback, \emph{(ii)} Bob needs to
  trust, and to have the same food taste as his feedback providers.

\item In all other cases, for example when Bob is abroad or when Sally's
  restaurant just opened, Bob uses a different mechanism. Indeed, Bob will rely
  on his own experience about restaurants to make a first estimate on Sally's offer: he
  will look at the global presentation of Sally and her restaurant, the amount
  of other clients, the menu, the prices and the location. Then he will compare
  this with what he knows about the other restaurants in the same city/country,
  of the same type, or where he previously had good/bad experience with the food
  he wants to order.
  This is essentially how our proposal works. Instead of relying on a chain of recommendation
  to assess trustworthiness, it relies on the user's (Bob) personal experience. Of
  course, the success of this assessment relies on two hypotheses: $(i)$ the
  user has enough experience, and $(ii)$ it is possible to learn and predict, from these
  experiences, the outcome of a new transaction. Our experimental results,
  conducted using data from an auction site show that these assumptions do hold in (some) real scenarios, where our proposal manages to predict efficiently auction frauds.
\end{itemize}

Note that our proposal is a generic trust framework, so various machine learning algorithms can be integrated, demonstrating that trustworthiness can be efficiently learned. In this work,
we use two common but effective machine learning algorithms: linear discriminant analysis (LDA) \cite{da, fukunaga1990} and decision tree (DT) \cite{introDT, mlintroduction} as the case studies for presentation, experiments and validation of our proposal. LDA is a well known method for dimensionality reduction and classification. It takes as input a set of events belonging to $k$ ($\geq$ 2) different classes and characterized by various features, and finds a
combination of the features (a classifier) that separates these $k$ classes of events.  As an example, linear discriminant analysis was used to differentiate subspecies of beetles based on measurements of their physical characteristics. DT is a widely used and practical method for classification and prediction by providing a classifier in the form of a tree structure. Its popularity is due to the ability of generating rules which can be easily expressed visually and in human language.

Although LDA and DT are simple, as seen in our real auction data driven experiments, they are both effective and efficient. Given proposed generic and extendable framework, more sophisticated methods (e.g., multiple discriminant analysis) may be applied. The associated trade-offs are not explored in this paper.

In our proposal, the agent's local knowledge (i.e., the past interactions) are described
by a set of features. Without loss of generality, we assume two classes:
successful and unsuccessful interactions. Note that all the
interactions have the same feature set. For LDA, features must be
quantitative, but can take values on an unbound domain. The agent who needs
to evaluate the potential interaction divides its historical interactions into
two groups: successful and unsuccessful. He then performs a LDA on this two
groups to obtain a linear classifier that allows him to estimate whether
the potential interaction is likely to get classified in the successful group or
not. For DT, the features can be quantitative or qualitative. A decision tree is firstly constructed from the training data. Then the algorithm classifies an example data by starting from the root of the tree and moving (down) until a leaf node, which is actually the classification of the example data. Since different machine learning algorithms may produce different results in the same scenario (due to their suitability), we also investigate confidence of each algorithm recommendation to help trustor make the wiser decision.
Please note that our proposal is designed for decentralized systems, since each agent uses its own local knowledge to evaluate trustworthiness of a transaction he might get involved in. However, this framework can be also applied in a centralized context, where all agents' local knowledge is gathered in a system-wide knowledge repository. In this context, multiple local knowledge can be aggregated to issue a better trust assessment.

In case trustor does not have sufficient local knowledge, we propose to construct a local knowledge sharing overlay network (LKSON), which allows local information (i.e., intermediate results of machine learning algorithms) to be exchanged between trustworthy agents. Such mechanism helps inexperienced agents to bootstrap in the system.

The contribution of this paper is to define a generic trust framework based on machine learning that is designed/suited for large-scale open systems. Since our proposal generally relies on machine learning algorithms (e.g., LDA and DT in this paper), we argue that trustworthiness of a potential transaction can be efficiently learned using trustor's local knowledge and the result of learning provides a good guidance on decision making. Moreover, confidence of the algorithm recommendations is studied to make decision wisely. Our proposal can be treated as a complimentary and alternative mechanism to provide prediction for a potential interaction, especially when there is no or few (reliable) global information available. Moreover, compared to existing reputation based trust models, our proposal is more robust: (1) it mainly relies on trustor's local knowledge thus lowering the risk of suffering inaccurate third party information; (2) features are difficult to fake since malicious agent is unaware of (or much efforts are needed to investigate) trustor's local knowledge thus may find it difficult to suitably modify features to mislead trustor. Our proposal also allows efficient information sharing to bootstrap inexperienced agents. Simulation based evaluation shows that compared to other trust models, our proposal is efficient, especially when third party information is not reliable. Moreover, performance of our proposal is quite stable because it only relies on trustor's local knowledge. Real auction dataset based evaluation also demonstrates the efficacy of our proposal in a realistic setting, e.g., for detecting auction frauds.

Note that since our proposal is particularly developed for evaluating trustworthiness of the transactions which are conducted by unknown service providers, we do not discuss how to maintain trust ratings of known agents, which can be easily handled by any traditional trust management mechanism.

The rest of this paper is organized as follows:
Generic trust framework is introduced in Section \ref{sec:trustframework}. In Section \ref{trustML}, we demonstrate how our proposal works using machine learning algorithms LDA and DT as the case studies in subsections \ref{sec:lda} and \ref{sec:dt} respectively. Confidence of algorithm recommendation is studied in Section \ref{sec:algoAccuracy}. We then discuss the issue of sharing local knowledge among the agents in Section \ref{sec:sharedmechanism}. Evaluation is performed on our proposal in Section \ref{sec:evaluation} using both real online auction dataset and synthetic dataset. In Section
\ref{sec:relatedwork} we present related works regarding trust management. Finally we conclude our work and suggest some future research directions in Section \ref{sec:conclusion}.

\section{A generic trust framework}
\label{sec:trustframework}

Unlike existing trust mechanisms, our proposal does not capture the behavior of
individual agent to decide whether or not to interact with it but focuses on a
potential interaction itself by estimating its reliability based on features of this interaction and trustor's local knowledge. This avoids
the risk of suffering inaccurate third party information.


\subsection{Notation}
\label{sec:notation}
We refer to a participant in the system as an agent. We denote by $\mathcal{A}$ the
set of all agents in the system. There are two types of agents, \emph{customers},
which request service from other agents and \emph{providers}, which provide
service. An agent can be both customer and service provider but in a specific transaction, it plays only one role. A transaction in the system happens when an agent accepts another agent's service. To indicate the quality of a service, an agent can rate the transaction. In our model, the rating of a transaction is binary: successful or unsuccessful.
$\Theta_{a_{x},a_{y}}$ denotes the transaction between provider agent $a_{y}$ and customer agent $a_{x}$. Note that without loss of generality, each $\Theta$ represents
a unique transaction. If multiple transactions happen between the same pair of
agents, we simply create a new virtual agent with the same characteristics.
Each transaction $\Theta$ is described by a set of features, denoted by
$F_{\Theta} = \{f^{1}_{\Theta}, f^{2}_{\Theta}, ..., f^{d}_{\Theta}\}$.

Each agent $a_{x}$ in the system maintains a transaction database $Tr_{a_{x}} = \{\Theta_{a_{x},a_{1}},
\Theta_{a_{x},a_{2}}, ...\}$, which records historical
transactions of this agent with other agents. Such local knowledge is used by
our trust framework to perform machine learning algorithm to predict trustworthiness of a potential transaction.

\subsection{Feature collection}
\label{sec:metainfocollection}
Features can be collected from profile of the transaction partner, or
the context of the transaction depending on specific applications. Let us take
online auction as an example to demonstrate how features are
collected. Buyer Bob is evaluating a potential transaction of purchasing a
camera from the seller Sally. The features associated with this
transaction can be collected from three aspects:
\begin{itemize}
	\item[(1)] about Sally herself, e.g., age, gender, is she from the same country as bob? physical distance to bob, etc.
	\item[(2)] about Sally as a user in the system, e.g., age in the system, numbers of successful and unsuccessful transactions, does Sally has provided a phone number? how complete is her profile? number of items Sally already sold, average delivering time, average time between end of auction and user comment, number of friends Sally has, etc.
	\item[(3)] the cameras Sally sells, e.g., average item price (items in the same category), number of the same items in stock, number of comments on these cameras, number of different buyers that already placed bid on it, average age in the system of buyers that already placed bid, etc.
\end{itemize}
Two points regarding feature collection need to be noted: (i) Sally can cheat and set fake information about herself, but faking information from the $2nd$ and $3rd$ aforementioned categories is harder since this information is maintained by the service exchange platform.
(ii) Although, if available and appropriate, our proposal also uses historical information (e.g., numbers of successful and unsuccessful transactions, etc.) like traditional trust mechanisms, the information is used in a quite different way, i.e., it is not directly used to derive trust but is used as a feature to help extract relationship between the potential transaction and past successful/unsuccessful transactions.
We will show in evaluation that by suitably using historical information as the features, our proposal outperforms traditional trust mechanisms that directly use historical information to derive trust.

We argue that compared to existing reputation based trust models, the feature collection process of our proposal is more robust against malicious
agents: for a traditional reputation based trust model, when trustor requests
opinions about the target agent, malicious entities can easily counterfeit the
feedbacks to promote or bad-mouth the target agent and trustor needs much
efforts to filter out these false feedbacks, but for our proposal, since attackers
are unaware of trustor's local knowledge, it is difficult to fake features associated with the potential transaction to mislead the trustor. Our proposal is thus more resilient to the attacks that are common to current reputation based trust systems.

\subsection{Architecture}
\label{sec:architecture}
\begin{figure}[ht]
\centering
\includegraphics[width=11.5cm]{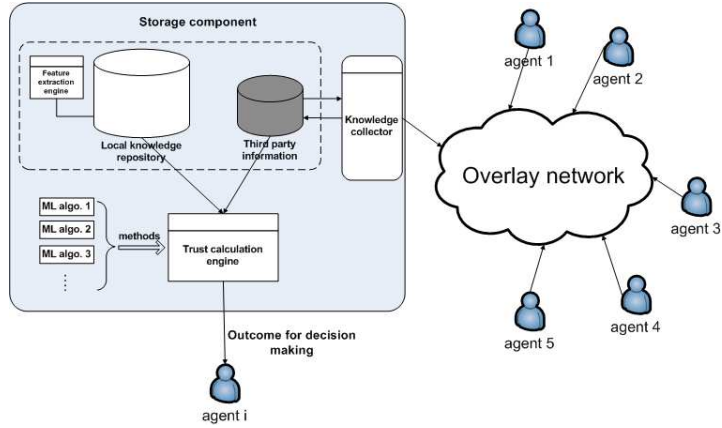}
\caption{The architecture.}
\label{fig:trustframework}
\end{figure}

Fig. \ref{fig:trustframework} depicts the architecture of our proposal. The framework consists of three components: storage component (SC), trust calculation engine (TCE) and knowledge collector (KC).

Storage component is responsible for managing local knowledge, including agent's past transactions as well as third party knowledge collected from other agents. After each transaction, agent records associated information such as characteristics of the transaction partner, outcome of the transaction and other related features (done by feature extraction engine), depending on the types of applications. For instance, in online auction site, such features may be price or category of the item, or number of items already sold by the seller, etc. (see Table \ref{tab:structureTD} as an example using Allegro \cite{allegro} dataset). These past transactions are used as the knowledge base to perform machine learning algorithms.

\begin{table}[!tb]
\centering
\caption{Structure of local knowledge repository}
\label{tab:structureTD}
\begin{tabular}{|l|p{2.2cm}|p{1.25cm}|p{1.5cm}|p{2cm}|p{0.5cm}|}
\hline
Tran. ID & Tran. partner & outcome & Price of the item & $\sharp$ of items already sold & ...\\
\hline
$\Theta_{1}$ & $a_{1}$ & 1 & 6.99  & 891 &  ... \\
\hline
$\Theta_{2}$ & $a_{2}$ & 0 & 9.99 & 337 &  ... \\
\hline
$\Theta_{3}$ & $a_{3}$ & 0 & 12.99 & 120 &  ... \\
\hline
...... &&&&&\\
\hline
\end{tabular}
\end{table}

Our trust framework is generic in two senses, the approach can be applied for different applications and correspondingly diverse features; and diverse machine learning approaches may be plugged in in the framework.
Trust calculation engine is responsible for applying machine learning algorithms to predict trustworthiness of the potential transaction using local knowledge or collected knowledge as the training set. Depending on application scenario, TCE is able to choose the appropriate machine learning algorithm to conduct trust calculation. For instance, if features are quantitative, discriminant analysis will be used, otherwise, if several features are qualitative, DT is more suitable. Actually, subject to time/computation restriction, TCE will apply as many suitable algorithms as possible. In case the result is not consistent across all algorithms, TCE returns the result that is calculated by the algorithm which has the highest confidence. Measurement of such confidence is algorithm dependent and we will discuss this issue in Section \ref{sec:algoAccuracy}.

With sufficient local knowledge and suitable machine learning algorithms, trustor is able reliably predict trustworthiness of a potential transaction. However, when local knowledge is insufficient, machine learning algorithms will perform poorly. To address this issue we
propose to construct a local knowledge sharing overlay network (LKSON) where agents are able to share their local information (done by knowledge collector (KC)). Different from traditional trust mechanisms where feedbacks of the specific agent are shared, in our approach, agents only exchange intermediate machine learning algorithm result (see Sec. \ref{sec:sharedmechanism}). Such a strategy has several advantages: (1) the shared information is just intermediate result of an algorithm so it is not easy to dig out agent's privacy, i.e., identification; (2) since information provider does not know whom the trustor is evaluating as well as trustor's local knowledge, it is difficult to send fake information to promote or bad-mouth some specific agent; (3) a lot of computation duplication is avoided. We will present the local knowledge sharing overlay network in detail in Section \ref{sec:sharedmechanism}. We next present details of our machine learning based trust assessment algorithms.

\section{Machine learning algorithms for trust assessment}
\label{trustML}
A machine learning algorithm for trust assessment takes as input a set of trustor's past transactions with other agents and the feature vector associated with each transaction. The output of the algorithm is a prediction indicating the potential transaction is reliable or risky, with a recommendation confidence (see local knowledge repository (with feature extraction engine) and trust calculation engine in Fig. \ref{fig:trustframework}).

We next use two common but effective machine learning algorithms, i.e., linear discriminant analysis and decision tree as the case studies to demonstrate how our proposal works. Algorithm recommendation confidence is discussed in Section \ref{sec:algoAccuracy}.

\subsection{Case study: Using linear discriminant analysis}
\label{sec:lda}
Consider a scenario where a customer $a_{x}$ encounters a
potential service provider $a_{y}$ and $a_{x}$ has no prior experience
with $a_{y}$. We assume that $a_{x}$ can obtain features of this
potential transaction $\Theta_{a_{x},a_{y}}$ (e.g., from $a_{y}$'s personal
profiles and from the transaction offer it received).  The set of such features of $\Theta_{a_{x},a_{y}}$ is denoted by $F_{\Theta_{a_{x},a_{y}}} =
\{f^{1}_{\Theta_{a_{x},a_{y}}}, f^{2}_{\Theta_{a_{x},a_{y}}}, ...,
f^{d}_{\Theta_{a_{x},a_{y}}}\}$.  (to simplify the notation, we
use $f^{i}$ instead of $f^{i}_{\Theta_{a_{x},a_{y}}}$ when the context is
clear).  So the potential transaction is represented by vector $p$ = ($f^{1}$
$f^{2}$ $f^{3}$ \ldots $f^{d}$).

We assume that $a_{x}$ has recorded $n$ historical transactions.
To estimate reliability of the potential transaction, $a_{x}$
classifies its historical transactions into two disjoint groups, the
\emph{successful transaction group} $G_{s}$ and the \emph{unsuccessful transaction group} $G_{u}$ according to
the outcomes of these transactions. The two transaction
groups are represented as\footnote{We assume an agent is aware of context of the transaction. For instance, when evaluating an unknown book seller, the buyer only considers past transactions regarding purchasing books, not that regarding purchasing other irrelevant items like digital cameras.}:

\begin{equation}
\mathbf{G_{s}} =
\left( \begin{array}{cccc}
\hspace{-0.2cm}f_{\Theta_{a_{x},a_{1}}}^{1}(s) & f_{\Theta_{a_{x},a_{1}}}^{2}(s) &...& f_{\Theta_{a_{x},a_{1}}}^{d}(s)\hspace{-0.2cm}\\
\hspace{-0.2cm}\vdots & \vdots & \vdots & \vdots \hspace{-0.2cm}\\
\hspace{-0.2cm}f_{\Theta_{a_{x},a_{n_s}}}^{1}(s) & f_{\Theta_{a_{x},a_{n_s}}}^{2}(s) &...& f_{\Theta_{a_{x},a_{n_s}}}^{d}(s)\hspace{-0.2cm}
\end{array} \right)
\end{equation}

\begin{equation}
\mathbf{G_{u}} =
\left( \begin{array}{cccc}
\hspace{-0.2cm}f_{\Theta_{a_{x},a_{1}}}^{1}(u) & f_{\Theta_{a_{x},a_{1}}}^{2}(u) &...& f_{\Theta_{a_{x},a_{1}}}^{d}(u)\hspace{-0.2cm}\\
\hspace{-0.2cm}\vdots & \vdots & \vdots & \vdots \hspace{-0.2cm}\\
\hspace{-0.2cm}f_{\Theta_{a_{x},a_{n_u}}}^{1}(u) & f_{\Theta_{a_{x},a_{n_u}}}^{2}(u) &...& f_{\Theta_{a_{x},a_{n_u}}}^{d}(u)\hspace{-0.2cm}
\end{array} \right)
\end{equation}

$n_{s}$ and $n_{u}$ are the sizes of successful transaction group and unsuccessful transaction group respectively. Please note that  $n = n_{s} + n_{u}$.

$a_{x}$ performs linear discriminant analysis to classify the potential
transaction as belonging to successful or unsuccessful transaction group to
decide whether or not to transact with the corresponding service provider for the specific transaction. The
mathematical operations are the following. Let $h_{x}$ be a $x\times 1$ (column)
vector of ones.

$a_{x}$ first calculates centroid of each group. That is, for each feature, we calculate average value of this feature across all transactions in the group.


\begin{equation}
    \label{eq:centroidS}
    \begin{split}
    c_{s} = \frac{1}{n_{s}}\cdot h_{n_s}^TG_s \textnormal{ and } c_{u} = \frac{1}{n_{u}}\cdot h_{n_u}^TG_u,
    \end{split}
\end{equation}



Similarly, the global centroid is calculated by averaging each feature across all past transactions:
\begin{equation}
    \label{eq:centroidAll}
    \begin{split}
      c=\frac{1}{n}\cdot h_{n}^{T}\left[
\begin{array}{c}
G_s  \\
G_u
\end{array}\right]
    \end{split}
\end{equation}

In LDA, the internal variance (within-class scatter matrix) and external variance (between-class scatter matrix) are used to indicate the degree of class separability, i.e., to what extent are the successful transactions distinguished from the unsuccessful transactions.
The internal variance, which is the expected covariance of each group is given by Eq. \ref{eq:withinS} and \ref{eq:withinU}:

\begin{equation}
    \label{eq:withinS}
    S_{w}^{s} = \frac{1}{n_{s}}(G_{s} - h_{n_s}c_{s})^T(G_{s} - h_{n_s}c_{s}),
  \end{equation}

  \begin{equation}
    \label{eq:withinU}
    S_{w}^{u} = \frac{1}{n_{u}}(G_{u} - h_{n_u}c_{u})^T(G_{u} - h_{n_u}c_{u}).
\end{equation}

$h_{n_s}$ and $h_{n_u}$ are used to center the values around the
centroids $c_{s}$ and $c_{u}$. So the overall within-class scatter matrix is calculated as the weighted sum of each group's internal variance, where the weight is fraction of transactions regarding the corresponding group:

\begin{equation}
    \label{eq:withinAll}
    S_{w} = \frac{1}{n}(n_{s}S_{w}^{s} + n_{u}S_{w}^{u}).
\end{equation}

Then $a_{x}$ calculates external variance, which is actually the covariance of the two groups, each of which is represented by its mean vector (Eq. \ref{eq:betweenModel}).

\begin{equation}
    \label{eq:betweenModel}
    S_{b} = \frac{1}{n}(n_{s}(c_{s} - c)^T(c_{s} - c) + n_{u}(c_{u} - c)^{T}(c_{u} - c)),
\end{equation}

The mixture scatter matrix, i.e., the total variance of the system is obtained by summing up internal variance $S_{w}$ and external variance $S_{b}$ (Eq. \ref{eq:mixtureModel}).

\begin{equation}
    \label{eq:mixtureModel}
    S_{m} = S_{w} + S_{b}.
\end{equation}

LDA aims to find a projection direction (a transformation) $v$ that maximizes the
inter class variance and minimizes the intra class variance. Formally, the criterion function (see Eq. \ref{eq:criterion}) is to be maximized.

\begin{equation}
   \label{eq:criterion}
   J(v) = \frac{v^{T}S_{b}v}{v^{T}S_{w}v}
\end{equation}

The projection direction $v$ is found as the eigenvector associated with the largest eigenvalue of $S_{w}^{-1}S_{b}$.
We then transform the two groups of transactions using $v$. Similarly, the potential
transaction $p$ = ($f^{1}$ $f^{2}$ $f^{3}$
\ldots $f^{d}$) is also transformed and classified by measuring the distances
between transformed potential transaction and the two groups (i.e., centroid). Eq. \ref{eq:distanceModelS}, \ref{eq:distanceModelU} show how to calculate such distances:
\begin{equation}
    \label{eq:distanceModelS}
    D_{s} = v^{T}p - v^{T}c_{s}
\end{equation}
\begin{equation}
    \label{eq:distanceModelU}
    D_{u} = v^{T}p - v^{T}c_{u}
\end{equation}

If $D_{u} > D_{s}$, then transaction $p$ is predicted as successful, otherwise it is predicted as unsuccessful.

Note that we try to collect as many features of a transaction as possible, and the algorithm filters out the not-so-relevant variables for us. That is to say, the feature which is more capable of distinguishing successful transactions from unsuccessful ones will have more impact on the final classification result.

\subsection{Case study: Using decision tree}
\label{sec:dt}
The basic idea of decision tree is to classify objects by sorting them from root of the tree to some leaf node, which provides classifications of the objects. Each node (except the leaf nodes) in the tree represents some feature of the object, and the value of such feature is the criteria by which the current set of objects are put in the final classifications or further split into smaller classes that will be classified according to other selected features (i.e., child nodes of current node). This process continues until all objects are classified.

When encountering a potential transaction from an unknown agent, trustor $a_{x}$ has $n$ past transactions with other agents, denoted by $Tr_{a_{x}} = \{\Theta_{1},
\Theta_{2}, ..., \Theta_{n}\}$. Each past transaction $\Theta$ is described by the same set of features $F_{\Theta} =
\{f^{1}_{\Theta}, f^{2}_{\Theta}, ..., f^{d}_{\Theta}\}$. Such knowledge is used to construct the decision tree to generate a classifier (rule) to predict whether a potential transaction is risky or not\footnote{Remind that we assume binary result of a transaction.}.

From a given past transaction set, it is possible to construct several different decision trees, depending on the order in which transaction features are tested. Since building an optimal decision tree (e.g., optimal with respect to its size) is NP hard \cite{dtNPhard} and therefore, several heuristics have been proposed to build efficient learning trees (e.g., ID3 \cite{introDT}, C4.5 \cite{c45}, etc.). Without loss of generality, we hereafter present one of these heuristics, namely ID3 \cite{introDT}. ID3 is a well known algorithm that relies on information gain \cite{informationtheory} to select the classifying features at each node of the tree. Information gain is measured by entropy.

Given the binary categorization (i.e., successful and unsuccessful) and past transactions, we denote the proportion of successful and unsuccessful transactions by $p_{s}$ and $p_{u}$ respectively. Then the entropy of all past transactions is:
\begin{equation}
   \label{eq:entropy}
   Entropy(Tr_{a_{x}}) = -p_{s}log_{2}(p_{s}) - p_{u}log_{2}(p_{u})
\end{equation}

Entropy is used to characterize (im)purity of a collection of examples. From Eq. \ref{eq:entropy} we can see entropy will have the minimum value of 0 when all past transactions belong to one class (i.e., successful or unsuccessful) and the maximum value of $log_{2}2 = 1$ when past transactions are evenly distributed across the two classes. Using entropy, we now calculate information gain of every feature (with respect to a set of past transactions $Tr$) to determine the best feature to choose for a node in the decision tree.
For each feature $f_{i}$, we assume it has a set of values (e.g., discrete variable) or intervals (e.g., continuous variable), which is denoted by $V(f_{i})$. For each $v \in V(f_{i})$, we denote the set of past transactions that are associated with $v$ for feature $f_{i}$ by $Tr_v$. The information gain of feature $f_{i}$ is thus calculated by:

\begin{equation}
    \label{eq:informationgain}
    IGain(Tr , f_{i}) = Entropy(Tr) - \sum_{v \in V(f_{i})}\frac{|Tr_{v}|}{|Tr|}Entropy(Tr_{v})
\end{equation}

The information gain of a feature measures expected reduction in entropy by considering this feature. Clearly, the higher the information gain, the lower the corresponding entropy becomes and thus the better the classification of past transactions is achieved (by considering the corresponding feature).

\begin{algorithm}[h]
\begin{algorithmic}[1]
   \STATE{Trust\_ID3($Tr_{a_{x}}$,$F$)}
       \IF{All past transactions are successful}
           \STATE{Return single node tree \emph{Root}, with label = +.}
       \ENDIF
       \IF{All past transactions are unsuccessful}
           \STATE{Return single node tree \emph{Root}, with label = --.}
       \ENDIF
       \IF{$F = \emptyset$}
           \STATE{Return single node tree \emph{Root}, with label = outcome of the transactions that dominate all past transactions.}
       \ENDIF
       \STATE{Calculate information gain of all features and choose one feature (say, $f$), which has the highest information gain as the \emph{Root} node.}
       \FOR{each value $v$ of $f$}
           \STATE Add a new branch below \emph{Root}.
           \STATE Let $Tr_{a_{x},f,v}$ be the set of past transactions that have value $v$ for feature $f$.
           \IF{$Tr_{a_{x},f,v} = \emptyset$}
               \STATE{Below this branch add a leaf node with label = outcome of the transactions that dominate all past transactions.}
           \ELSE
               \STATE{Below this branch add a sub-tree:}
               \STATE{Trust\_ID3($Tr_{a_{x},f,v}$,$F-\{f\}$).}
           \ENDIF
       \ENDFOR
   \STATE{\textbf{Return} \emph{Root}.}
\end{algorithmic}
\caption{Decision tree construction (code for trustor $a_{x}$).} \label{alg:id3} \end{algorithm}

Using information gain, a decision tree is iteratively constructed by maximizing information gain at each step. We next describe how decision tree is constructed (Algo. \ref{alg:id3}) by ID3 algorithm.

The ID3 algorithm input is trustor $a_{x}$'s past transactions $Tr_{a_{x}}$ and the feature vector $F$ that describe each transaction. If all transactions are successful or unsuccessful, the algorithm simply returns a single node (i.e., root) tree, labeled as '+' or '--' respectively (Line 2-7). If feature vector is empty, the algorithm returns a single node tree, labeled as '+' if most of past transactions are successful or '--' otherwise (Line 8-10). Then the algorithm calculates information gain of each feature using Eq. \ref{eq:informationgain}. The feature (say, $f$) with the highest information is selected as the root of the tree. Next, new branches are added to the root node according to possible values of $f$. For each possible value $v$, we denote the set of past transactions that have $v$ for $f$ by $Tr_{a_{x},f,v}$. If $Tr_{a_{x},f,v} = \emptyset$, the algorithm adds a leaf node to the corresponding branch with label = '+' or '--' if most of past transactions are successful or unsuccessful respectively. Otherwise, the algorithm recursively executes itself with $Tr_{a_{x},f,v}$ and $F-\{f\}$ as the input parameters (Line 12-21). The algorithm terminates either when all the features have been tested or the tree perfectly classifies the past transactions.

When encountering a potential transaction, $a_{x}$ tries to classify it as a risky transaction or not by comparing the corresponding features from root of the tree down to a certain leaf node. Note that for simplicity, we do not discuss other issues in decision tree such as avoiding overfitting the data, etc.

\section{Algorithm recommendation confidence}
\label{sec:algoAccuracy}

Given a features vector, a set of past transactions and appropriate machine learning algorithm, our trust framework is able to return a recommendation of the potential transaction: be it risky or not.
Confidence of such recommendation relies on characteristics of input (e.g., discriminating power of the features, volume of past transactions) and performance of the specific machine learning algorithms in use\footnote{Given the same input, different machine learning algorithms may generate different recommendations (perhaps due to their suitability in different application scenarios).}. So by performing the same algorithm, trustors with different local knowledge and/or feature vectors may have quite different confidence on the recommendations. We denote $\sigma \in [0,1]$ as the algorithm recommendation confidence, where 0 represents the algorithm recommendation is not confident at all (i.e., the recommendation is equally random) and 1 represents the algorithm recommendation is completely confident. Algorithm input is denoted by $<F,Tr>$, where $F = \{f_{1}, f_{2}, ...\}$ represents the feature vector and $Tr = Tr_{s}\cup Tr_{u}$ represents the set of past transactions: $Tr_{s} = \{\theta_{1}^{s}, \theta_{2}^{s}, ...\}$ and $Tr_{u} = \{\theta_{1}^{u}, \theta_{2}^{u}, ...\}$ are the sets of successful transactions and unsuccessful transactions respectively.
We next present that given different inputs, how to measure confidence of the algorithm recommendations still using linear discriminant analysis and decision tree as the demonstration examples.

\subsection{Measuring recommendation confidence for LDA based algorithm}
\label{sec:confidenceLDA}
According to theory of linear discriminant analysis based algorithm (refer to Section \ref{sec:lda}), each transaction is described by the input feature vector $F$ (i.e., each transaction is represented as a point in $|F|$-dimensional space.). So confidence of the algorithm recommendation greatly depends on discriminating power of the input features. If all the features are discriminating enough, LDA is able to separate the successful and unsuccessful transactions clearly and minimize the variance of each transaction group. Otherwise, the two transaction groups $Tr_{s}$ and $Tr_{u}$ may overlap, thus making classification of the potential transaction less confident. In other words, algorithm input determines the LDA projection direction. After having the project direction, transaction group centroids are transformed to lower dimension space. The potential transaction (after transformed) is classified as risky or not by measuring the distances between it and the two transaction group centroids. We denote the distances between the potential transaction and successful/unsuccessful transaction group by $D_{s}$/$D_{u}$. Note that each set of distances correspond to one specific input: $<D_{s},D_{u}>$ = LDA($<F,Tr>$), where LDA() performs linear discriminant analysis to generate distances using the given input. The confidence of the algorithm recommendation is then calculated as:

\begin{equation}
\label{eq:confidenceLDAEQ}
		\sigma_{LDA} = \frac{|D_{s} - D_{u}|}{D_{s} + D_{u}}
\end{equation}

It is clear from Eq. \ref{eq:confidenceLDAEQ} that when $D_{s}$ = $D_{u}$, $\sigma_{LDA} = 0$, i.e., the potential transaction is classified as risky or not with the same probability. When $D_{s}$ or $D_{u}$ is 0, $\sigma_{LDA} = 1$, i.e., the potential transaction is recommended to be safe or not confidently. With $\sigma_{LDA}$, trustor is able to measure how confident the algorithm outcome is and personally decides whether or not to trust the algorithm recommendation.

\subsection{Measuring recommendation confidence for DT based algorithm}
\label{sec:confidenceDT}

Since decision tree is constructed according to information gain of each feature (refer to Section \ref{sec:dt}), it is natural to measure algorithm recommendation confidence using feature's information gain. According to information theory, information gain of a feature measures expected reduction in entropy by considering this feature. If a feature is discriminating enough, the corresponding information gain is high (maximum value is $log_{2}2 = 1$), i.e., by appropriately choosing values of this feature, the associated successful transactions and unsuccessful transactions are separated clearly, thus the potential transaction can be accurately classified by this feature. Otherwise, this feature may not help classify the potential transaction accurately. So with different input data (i.e., feature vector and set of past transactions), different information gains for a specific feature may be generated and hence different confidence on the classification results may be obtained.

For each feature $f_{i}$ (with respect to a set of transactions $Tr_{i}$), we calculate the corresponding entropy $Entropy(Tr_{i})$ and information gain $IGain(Tr_{i} , f_{i})$ using Eq. \ref{eq:entropy} and \ref{eq:informationgain} respectively. We then calculate the extent of entropy reduction by feature $f_{i}$:

\begin{equation}
\label{eq:entropyratio}
		r_{i} = \frac{IGain(Tr_{i} , f_{i})}{Entropy(Tr_{i})}
\end{equation}

$r_{i}$ can be interpreted as the accuracy of classification of the potential transaction by feature $f_{i}$ (i.e., at the node corresponding $f_{i}$ in the decision tree.). We notice from Eq. \ref{eq:entropyratio} that if $f_{i}$ is extremely discriminating, its information gain is $Entropy(Tr_{i})$ (i.e., the second part of Eq. \ref{eq:informationgain} is 0) and thus $r_{i}$ is 1, which means the potential transaction classification by $f_{i}$ is completely accurate. If feature $f_{i}$ is incapable of discriminating, its information gain is 0 and thus $r_{i}$ is 0, which means classification of the potential transaction is inaccurate at all (i.e., the classification is equally random.).

After having accuracy of classification of each feature, we calculate algorithm recommendation confidence as:

\begin{equation}
\label{eq:confidenceDTEQ}
    \sigma_{DT} = \prod_{i = 1}r_{i}
\end{equation}

So, if all the features are quite discriminating, the algorithm recommendation confidence is approaching 1. Otherwise, any less discriminating feature shall greatly impact the final confidence. This also follows the decision rule of the decision tree based algorithm.

\section{Local Knowledge Sharing Mechanism}
\label{sec:sharedmechanism}
Our trust framework works under the assumption that the trustor has sufficient local knowledge such that machine learning algorithms can be performed. However, in some cases, agents may have no or sparse local knowledge (e.g., the bootstrapping agents recently join the system or do not interact with others frequently). To make our proposal adaptive to such a scenario, we propose to construct a overlay network dedicated to exchange/share agents' local knowledge to help inexperienced agents estimate trustworthiness of a potential transaction.

\subsection{Local Knowledge Sharing Overlay Network (LKSON)}
\label{sec:overlay}
\begin{figure}[h]
\centerline{\includegraphics[scale=0.65]{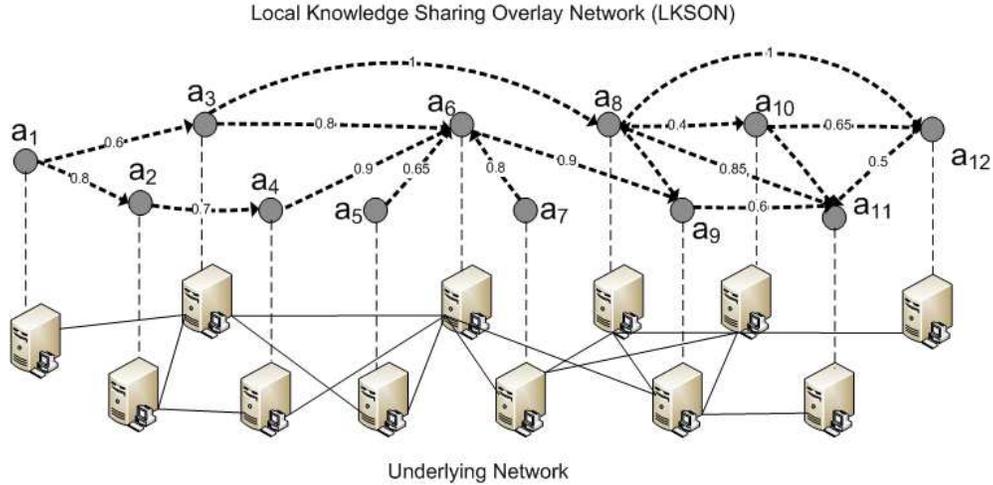}}
\caption{A local knowledge sharing overlay network.} \label{fig:overlay}
\end{figure}

To evaluate trustworthiness of a potential transaction, inexperienced agents have to request other agents' knowledge. A local knowledge sharing overlay network (LKSON), which is a virtual network on top of current network infrastructure (e.g., P2P network) is thus constructed. Fig. \ref{fig:overlay} depicts a LKSON, which is represented by a directed graph. We give definition of LKSON:

\begin{definition}[Local Knowledge Sharing Overlay Network]
A local knowledge sharing overlay network (LKSON) is a Weighted Directed Network (WDN), denoted by LKSON = (V,E,T):\\
    $V = \{a_{1}, a_{2}, a_{3}, ...\}$ is the set of agents in LKSON.\\
    $E = \{e_{i,j}|a_{i}\in V \bigwedge a_{j}\in V \bigwedge a_{i}\neq a_{j}\}$ is the set of directed edges representing the trust relationship in terms of providing correct local knowledge between knowledge requester $a_{i}$ and requestee $a_{j}$.\\
    $T:E \rightarrow [0,1]$, trust score $t_{i,j}$ is assigned to edge $e_{i,j}$ and expresses from viewpoint of $a_{i}$, how trustworthy $a_{j}$ is in terms of providing local knowledge.
\end{definition}

We set that the trust score falls into the range of [0,1], where 1 represents completely trustworthy and 0 represents completely untrustworthy. Trust score is determined according to requestee's past performance (i.e., quality of provided local knowledge). After each interaction, knowledge requester $a_{i}$ evaluates usefulness of the information provided by the requestee $a_{j}$ and accordingly updates trust score of $a_{j}$. Please note that since agent may have very personalized perspective on the system, even if requestee $a_{j}$ provides correct local knowledge, it may not be useful to the requester. In this situation, $a_{j}$'s trust score will be lowered although it is honest. We will discuss details of trust score update in Section \ref{sec:LKSONupdate}.

Notice that LKSON is constructed to ensure reliable local knowledge sharing to help inexperienced agents estimate trustworthiness of a potential transaction, but is not used to discover trustworthy transaction partner (i.e., LKSON is not a trust overlay network like that in \cite{powertrust}) because it has been reported in related literature \cite{Xiong04peertrust}, \cite{Swamynathan05decouplingservice} that the agents who provide high quality service may not report genuine feedbacks and vice versa due to various reasons. In the case that the agents who provide useful information also act honestly in a transaction, LKSON can be used to help promote successful transactions (i.e., selecting reliable agents who have high trust scores as the service providers). However, the goal of this work is to design mechanisms to estimate trustworthiness of a potential transaction, which is conducted by an unknown agent (i.e., no historical information is available), so discussion on relying LKSON to estimate agent's trustworthiness like traditional trust mechanisms (e.g., feedbacks aggregation \cite{betareputation02}, \cite{p2prep}, \cite{travos06}, web of trust \cite{distributedtrust97} \cite{transitive03}, etc.) is out of the scope of this paper.

In order to collect useful information, inexperienced agents only request the agents who are trustworthy (i.e., trust score is high) in terms of providing useful local knowledge. So in our approach, each agent $a'$ in the system maintains a \emph{Trusted Knowledge Provider} list $PR_{a'}$ which stores the agents that this agent has trust relationship in terms of providing useful local knowledge.
Initially, when requestees' past performance is not available, an inexperienced agent requests information by exploring the social network. For instance, it may request
its ``familiar'' agents (e.g., friends or colleagues in the real world, etc.); or the agent may request ``special'' agent/entity (if any) in the network (e.g., super peer in hybrid peer-to-peer network, or advisor in Epinions \cite{epinions}, or authorized third parties, e.g., ``official'' source in public key infrastructure, etc.). In the case that no ``familiar'' or ``special'' agents are there (i.e., no useful information at all), the inexperienced agent will choose knowledge provider randomly to create its \emph{Trusted Knowledge Provider} list.

We next present detail of LKSON design using LDA as the demonstration example. Some issues like what kind of information is shared, how to combine knowledge of multiple sources and how to update relationships among agents in LKSON, etc. will be discussed.

\subsubsection{What is Shared?}
\label{sec:overlaywhatisshared}
To help an inexperienced agent detect a potentially risky transaction, two types of local knowledge, which are represented as a tuple $<$$v$, ($c_s$, $c_u$)$>$ are shared among the agents\footnote{We assume that all the agents have the same features such that the projection direction $v$ can be shared and combined meaningfully.}:
\begin{itemize}
    \item[(i)] the projection direction $v$ that maximizes
        the inter class variance and minimizes the intra class variance (see Sec. \ref{sec:lda}).
    \item[(ii)] centroid of the successful group $c_{s}$ and unsuccessful group $c_u$.
\end{itemize}
The main reasons that these two types of knowledge are shared: (1) the knowledge providers have no motivation to send fake information because they are unaware of requester's potential transaction partner thus are unable to defame/promote the transaction partner or mislead the requester maliciously\footnote{This reason is sometimes not true (e.g., the knowledge provider may learn requester's transaction partner from other ways, or the provider simply sends incorrect information for fun, etc.). We address this issue by updating \emph{Trusted Knowledge Provider} list, which will be described in Sec. \ref{sec:LKSONupdate}.}. That is, this is safer than directly requesting others' trusts in the target agent, which is very easy to be attacked). (2) the projection direction $v$ is shared, which means requester does not need to compute it so a lot of computation is saved. The process of deriving projection direction is the most computation-intensive part of LDA.
(3) the shared knowledge is just projection direction and centroid of successful and unsuccessful groups so it is not easy to dig out agent's privacy through shared information. This is important to knowledge providers and their past transaction partners. (4) The shared knowledge is of small size so the communication overheads incurred are marginal compared
to the cost of sending the whole local knowledge base.

\subsubsection{Combining Third Party Knowledge}
\label{sec:combiningknowledge}
When an agent $a_{i}$ (with no local knowledge) encounters a potential transaction $p = (f^1 f^2 f^3 ... f^d)$, where $f$ represents feature. $a_{i}$ requests local knowledge of other agents that are recorded in $a_{i}$'s \emph{Trusted Knowledge Provider} list $PROVIDER(a_{i})$ to predict trustworthiness of $p$. We denote the set of knowledge providers who response and provide the requested information by $KP_{i} = \{a_{1}^{i}, a_{2}^{i}, a_{3}^{i}, ..., a_{n_{i}}^{i}\}$. So the size of $KP_{i}$ is $n_{i}$. The corresponding trust scores of these knowledge providers are denoted by $TS = \{t_{1}^{i}, t_{2}^{i}, t_{3}^{i}, ..., t_{n_{i}}^{i}\}$, where $t_{j}^{i}$ represents $a_{i}$'s trust in the $j$th knowledge provider. The local knowledge shared by the $j$th knowledge provider is represented by $<$$v_{j}$, ($c_{s,j}$, $c_{u,j}$)$>$.

The collected knowledge is combined to produce a classifier that helps estimate whether $p$ is likely to get classified to the successful group or not. The resultant projection direction is derived as the weighted sum of all collected projection direction:
\begin{equation}
    \label{eq:projectiondirection}
    v_{i} = \sum_{j}^{n_{i}}w_{j}v_{j}
\end{equation}

Similarly, the resultant centroid of successful group and unsuccessful group are calculated as:
\begin{equation}
\label{eq:resultantcentroidS}
    c_{s,i} = \sum_{j}^{n_{i}}w_{j}c_{s,j}
\end{equation}
\begin{equation}
\label{eq:resultantcentroidU}
    c_{u,i} = \sum_{j}^{n_{i}}w_{j}c_{u,j}
\end{equation}

The weight $w_{j}$ for the $j$th knowledge provider is determined by its trust score:
\begin{equation}
    \label{eq:weight}
    w_{j} = \frac{t_{j}^{i}}{\sum_{j}^{n_{i}}t_{j}^{i}}
\end{equation}

After combining other knowledge providers' projection direction ($v_{i}$), $a_{i}$ transforms the combined centroid ($c_{s,i}$, $c_{u,i}$) using $v_{i}$. Similarly, the potential transaction $p$ is also transformed and classified using Euclidean distance of transformed data from each transformed centroid. The distances are computed using Eq. \ref{eq:distanceModelS} and \ref{eq:distanceModelU}.

\subsection{LKSON Update}
\label{sec:LKSONupdate}
Since behaviors of the agents may change overtime (e.g., honest knowledge provider may provide fake information or they may change their perspective on the system thus making the shared knowledge not useful to the requester any more, etc.), it is essential to update trust scores of the knowledge providers such that the inaccurate/useless knowledge has less impact on the final decision.

It is effective if we adapt trust scores of the knowledge providers accordingly every time a transaction is completed. However, such strategy is inefficient because it incurs unnecessary computation overheads. That is, only trust scores of the knowledge providers whose local knowledge produces contrary prediction on the outcome of the transaction need to be updated. So we adapt a strategy by trading off computation overhead and performance, that is, after one transaction, if its outcome is unsuccessful, the requester $a_{i}$ evaluates usefulness of local knowledge of each knowledge provider. For knowledge provider $a_{j}^{i}$, $a_{i}$ simply uses $v_{j}$ as the projection direction to calculate distances of transformed potential transaction from the transformed centroid (see Eq. \ref{eq:distanceModelS} and \ref{eq:distanceModelU}). If the transaction is predicted as successful, it means the corresponding knowledge is incorrect or useless to $a_{i}$ so trust score of the knowledge provider $a_{j}^{i}$ must be lowered such that it has less impact on future decision. Otherwise, the trust score should be raised. The trust score update follows a Bayesian approach \cite{betareputation02}, which takes binary ratings as input and is based on computing trust scores by statistically updating beta probability density function (PDF). The posteriori trust score is computed by combining a priori trust score with new evidence. For knowledge provider $a_{j}^{i}$, we denote numbers of its successful and unsuccessful predictions by $s_{j}^{i}$ and $u_{j}^{i}$ respectively (i.e., the original trust score $t_{j}^{i}$ is $\frac{s_{j}^{i} + 1}{s_{j}^{i} + u_{j}^{i} + 2}$). So $a_{j}^{i}$'s trust score is updated after one transaction using Eq. \ref{eq:trustupdateS} (the new prediction is correct) and Eq. \ref{eq:trustupdateU} (the new prediction is incorrect).

\begin{equation}
\label{eq:trustupdateS}
    t_{j}^{i} = \frac{s_{j}^{i} + 2}{s_{j}^{i} + u_{j}^{i} + 3}
\end{equation}

\begin{equation}
\label{eq:trustupdateU}
    t_{j}^{i} = \frac{s_{j}^{i} + 1}{s_{j}^{i} + u_{j}^{i} + 3}
\end{equation}

On the other hand, if the transaction is successful, the requester will evaluate usefulness of a small fraction (e.g., 20\%) of the third party knowledge. This can help (1) avoid incurring too much unnecessary computation overheads because if the transaction is successful, it means combination of current third party knowledge is accurate enough to make a good prediction so no update is needed urgently and (2) proactively adjust weights of the third party knowledge thus making the combined knowledge more accurate (for the next prediction) while keeping the computation overheads low.

\section{Evaluation}
\label{sec:evaluation}

\subsection{Simulation settings}
\label{sec:setting}

We use real dataset collected from an Internet auction site
Allegro \cite{allegro} as well as synthetic data to conduct
experiments. The Allegro dataset contains 10,000 sellers, 10,000 buyers, more
than 200,000 transactions and over 1.7 million comments. In the experiments, a
transaction is considered successful if its feedback is positive, otherwise, it
is considered unsuccessful. We extract three features from Allegro data, i.e., $F_{1}$:
category of the item; $F_{2}$: price of the item, $F_{3}$: number of items
already sold by the seller when the transaction occurs and $F_{4}$: fraction of non-positive feedbacks of the seller. We evaluate performance
of our proposal, i.e., LDA based approach and DT based approach by studying their capabilities of detecting Internet auction fraud. When a buyer encounters a potential transaction, which is conducted by an unknown seller (i.e., no historical information of this seller is available),
this buyer will gather past transactions regarding the item category (i.e., $F_{1}$) and then perform LDA or DT to estimate trustworthiness of this transaction using features $F_{2}$, $F_{3}$ and $F_{4}$.

Allegro has a very friendly environment: only around 0.9\% of the transactions
are labeled negative explicitly. To test our proposal in a more hostile
environment, and also to compare proposed algorithms with other existing trust mechanisms, we
generate synthetic dataset consisting of 10,000 nodes. Each node is either \emph{good}, which provides service satisfactorily
or gives genuine feedbacks when requested, or \emph{malicious}, which cheats
others in transactions or provides false feedbacks when requested.
To make the simulation environment more realistic, there are no completely \emph{good} or
\emph{bad} nodes. We set \emph{good} nodes and \emph{malicious} nodes behave
maliciously with probabilities of 15\% and 85\% respectively.
We denote the fraction of malicious nodes in the system by $P_{m}$.
In the simulation, any node (trustor) may
request service from other nodes (service provider or the target node). Once the transaction with service provider (target node) is done, trustor rates the transactions to indicate whether they are successful or not. The trustor stores information of these
transactions (see Table \ref{tab:structureTD}) such as their outcomes and the
associated features. In synthetic dataset, to construct LKSON, each node also maintains a \emph{Trusted Knowledge Provider} list, which records other nodes that will be requested if this node does not have sufficient local knowledge.

We also need to generate synthetic features. Note that each transaction is described by four features\footnote{We also conducted experiments using other numbers of features and obtained the similar qualitative trends.}. Let $X_s$ and $X_u$ be the
random variables which represent values of features of successful and
unsuccessful transactions respectively. We assume that $X_{s}$ and $X_{u}$
follow normal distribution, i.e., $X_s \sim \mathcal{N}(\mu,\sigma)$ and $X_u
\sim \mathcal{N}(\mu + \theta,\sigma)$. In our experiments $\mu=1$, $\sigma =
0.1$ and $\theta \in [0,1]$ is the separability factor that we use to tune the
overlap between values of features of successful transactions and that
of unsuccessful ones. When $\theta = 0$, values of features for
successful and unsuccessful transactions overlap completely, while when $\theta$ is large enough, it
means values of features for successful and unsuccessful transactions
are separated clearly. In other words, $\theta$ allows us to modify the
learnability of the transaction classes. Note that since only our approaches are based
on features, $\theta$ only impacts performance of our approaches.

We compare our proposal with three existing models: (1) \emph{Random Selection}: The simplest approach is to engage or reject a potential transaction randomly. In the simulation, we set the probability that trustor accepts a potential transaction to 50\%. This model is used as the experiments benchmark. (2) \emph{Feedback Aggregation}: In this model \cite{betareputation02, Xiong04peertrust, travos06}, if trustor does not know target node, it asks other nodes across the network and aggregate feedbacks to derive target node's trust. False feedback filtering mechanism (from TRAVOS \cite{travos06}) is applied to improve performance  (3) \emph{StereoTrust} \cite{stereotrust09}: Trustor forms groups of agents with which it has past experience according to the features and derives trust rating of the target agent by combining group trusts to which the target agent belongs. Please refer to related work section for a brief summary of StereoTrust.

Note that we only compare our proposal with the existing models when synthetic data is used because we lack the information of how users in Allegro apply these existing models.

The metrics we use to evaluate performance of trust mechanisms include:
\begin{itemize}
   \item \emph{false positive rate}\\
      The transaction is unsuccessful but the algorithm predicted that it would be successful.
   \item \emph{false negative rate}\\
      The transaction is successful but the algorithm predicted that it would be risky.
   \item \emph{overall falseness}\\
      Sum of false positive rate and false negative rate.
\end{itemize}

Each experiment is repeated 10 times, and error bars are added indicate deviation of each running.

\subsection{Results}
\label{sec:results}

\subsubsection{Real dateset}
\label{sec:realdatasetresult}
We rank the 10,000 buyers according to number of their past transactions, i.e.,
the first buyer has the most past transactions and the last one has the least
transactions. We select subset $U_{b}$ of these buyers starting from the first
one. Each buyer evaluates 100 randomly selected transactions (50\% are
successful and 50\% are unsuccessful). We vary the size of $U_{b}$ to investigate
effect of local knowledge volume.

\begin{figure}[tbp]
  \begin{center}
  \hspace*{-2.2cm}
  \centering
    \subfigure[\label{fig:varyingbuyersreal}Performance of our approach with varying number of buyers.]{\includegraphics[scale=0.232]{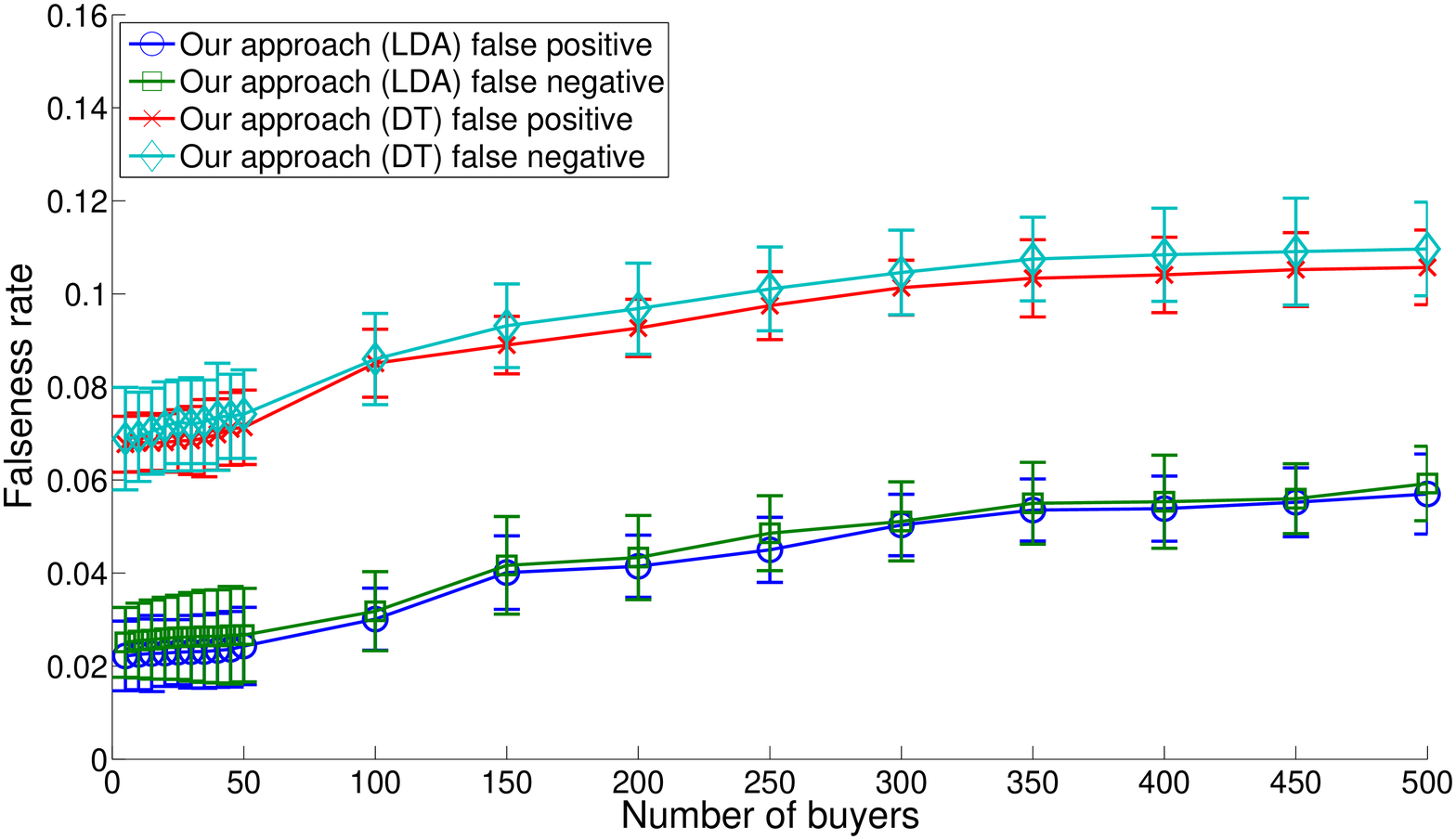}}\hspace*{-0.07cm}
    \subfigure[\label{fig:buyers}Numbers of individual buyers' past transaction.]{\includegraphics[scale=0.232]{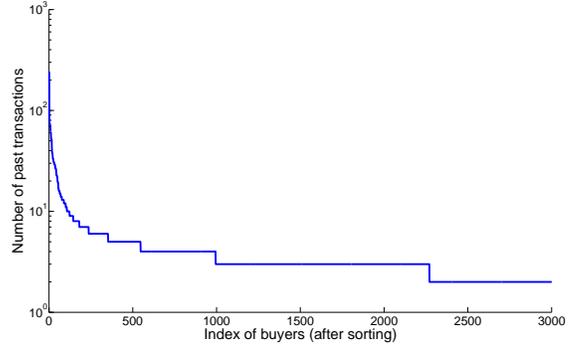}}\hspace*{-0.7cm}
  \end{center}
  \caption{Experiments using Allegro dataset.}
\label{fig:comparison04}
\end{figure}

Fig. \ref{fig:varyingbuyersreal} demonstrates performance of our approach (LDA and DT based), i.e., how average rates of false positive and false negative evolve when $U_b$ varies from 5 to 500. As expected, all falseness rates increase when $U_{b}$ grows.
This shows the impact of local knowledge on our proposed algorithms: when $U_b$ is small, it
contains only experienced agents, that all have enough past transactions to allow
our approach to issue accurate predictions. As $U_b$ grows, it contains more and
more inexperienced agents, for which the predictions are less accurate. We also notice that DT based algorithm is less accurate than LDA based algorithm. This is because basic decision tree is more suitable for qualitative features while features identified in auction data are quantitative (i.e., it is not easy to obtain the appropriate thresholds to efficiently split the training data.).

Fig. \ref{fig:buyers} shows the distribution of numbers of individual buyers'
past transactions (only first 3000 are shown). Note the logarithmic scale for
y-axis: the number of past transactions is quickly decreasing. For instance,
less than 100 agents have more than 20 past transactions. Estimating the minimal
number of transactions that allow our approach to be precise is challenging, since
not all transactions have the same importance (e.g., two very different transactions
will be much more useful for our approach than two identical transactions).
However, in this set of experiments, we estimate empirically that when numbers
of transactions is over 6, the potential transaction can be relatively reliably
predicted (i.e., the overall falseness rate is smaller than 0.1 for LDA based algorithm and 0.19 for DT based algorithm). Simulation
results thus show that our approach is capable of quickly ``learning'' what a
successful or unsuccessful transaction is.

In online auction sites, most of the buyers are inexperienced (local knowledge is insufficient). So to predict trustworthiness of the potential transaction, the inexperienced buyer will collect other buyers' local knowledge to perform our approach by constructing LKSON. LKSON is initialized by connecting this buyer and other buyers according to categories that they are interested in (i.e., the item this buyer is going to buy is in the category from which the connected buyers used to buy items). Third party knowledge indeed helps inexperienced buyer predict trustworthiness of the potential transaction, however, it also introduced inaccurate information, which may influence accuracy of prediction because third party knowledge only reflects knowledge provider's personalized perspective on the system. For instance, one buyer can tolerate one day delay of item delivery (successful transaction) while another buyer can not (unsuccessful transaction). We thus need to estimate how many other buyers are requested to gather sufficient knowledge while trying to keep influence of inaccurate information as low as possible.

\begin{figure}[h]
  \centering
  \includegraphics[scale=0.28]{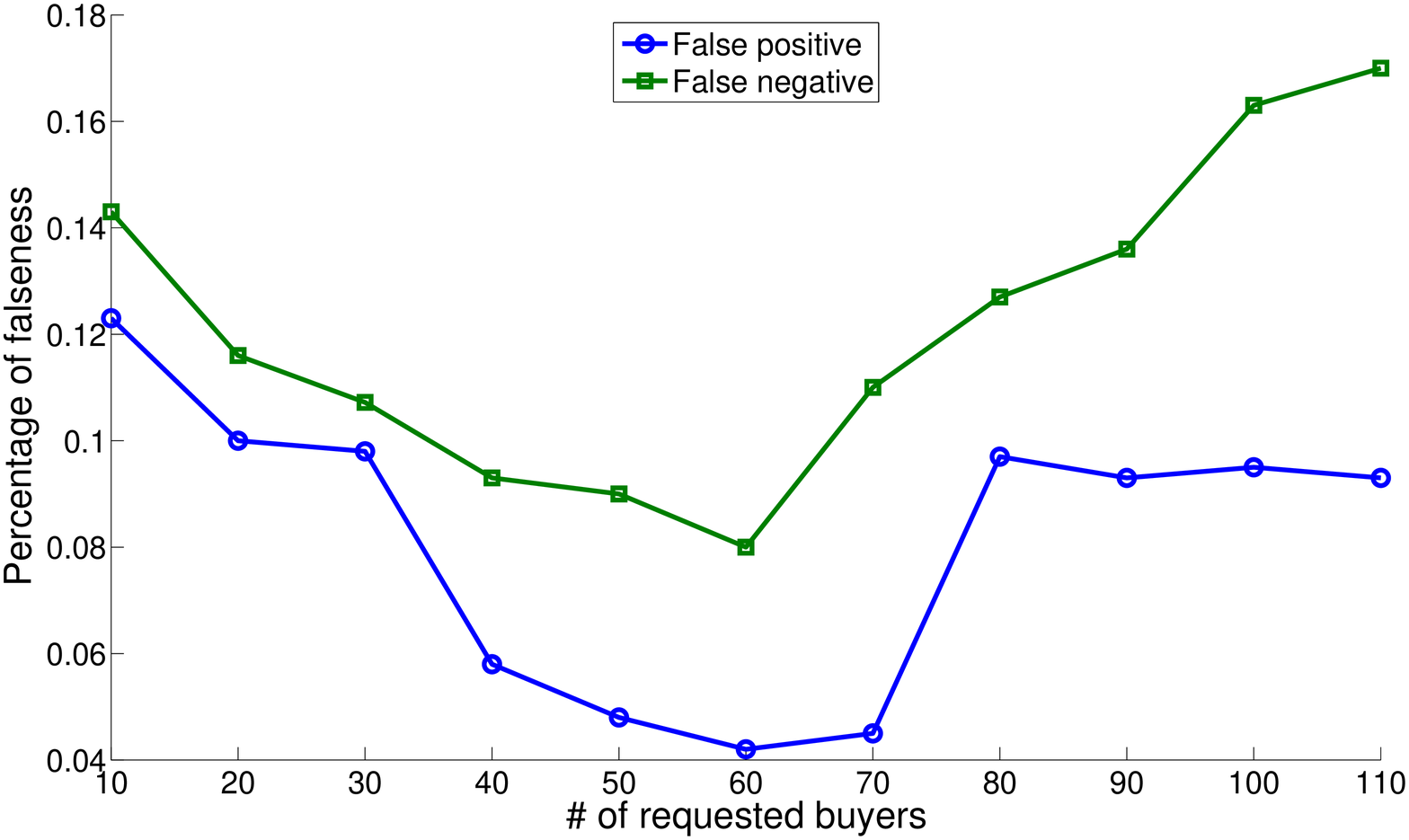}
    \caption{Performance of our approach (LDA) with LKSON (real dataset).}
  \label{fig:realwithLNSONFPFN}
\end{figure}

Fig. \ref{fig:realwithLNSONFPFN} demonstrates how performance of our approach (LDA based) varies with the varying number of requested buyers (from 10 to 110 with 10 as increment). We observe that the general trends of the false positive rate and false negative rate are: they decline with the increasing number of requested buyers and stop at a certain point, then they start ascending. This is because when the number of requested buyers is low, there are not sufficient past transactions available thus affecting the accuracy of prediction (i.e., not statistically significant). On the other hand, when the number of requested buyers is high, lots of transactions are collected, which means more inaccurate information is incorporated, hence the accuracy of prediction is also influenced. So the lowest point is the optimal result that our approach is able to achieve and the corresponding number of requested buyers is the design parameter we are looking for. From the figure we can see the most suitable number of requested buyers in this scenario is 60 and the corresponding false positive rate and false negative rate are 0.08 and 0.042 respectively.

\subsubsection{Synthetic dataset}
\label{sec:syndatasetresult}

\begin{figure}[tbp]
  \begin{center}
  \hspace*{-3.2cm}
  \centering
    \subfigure[\label{fig:FP02}False positive.]{\includegraphics[scale=0.18]{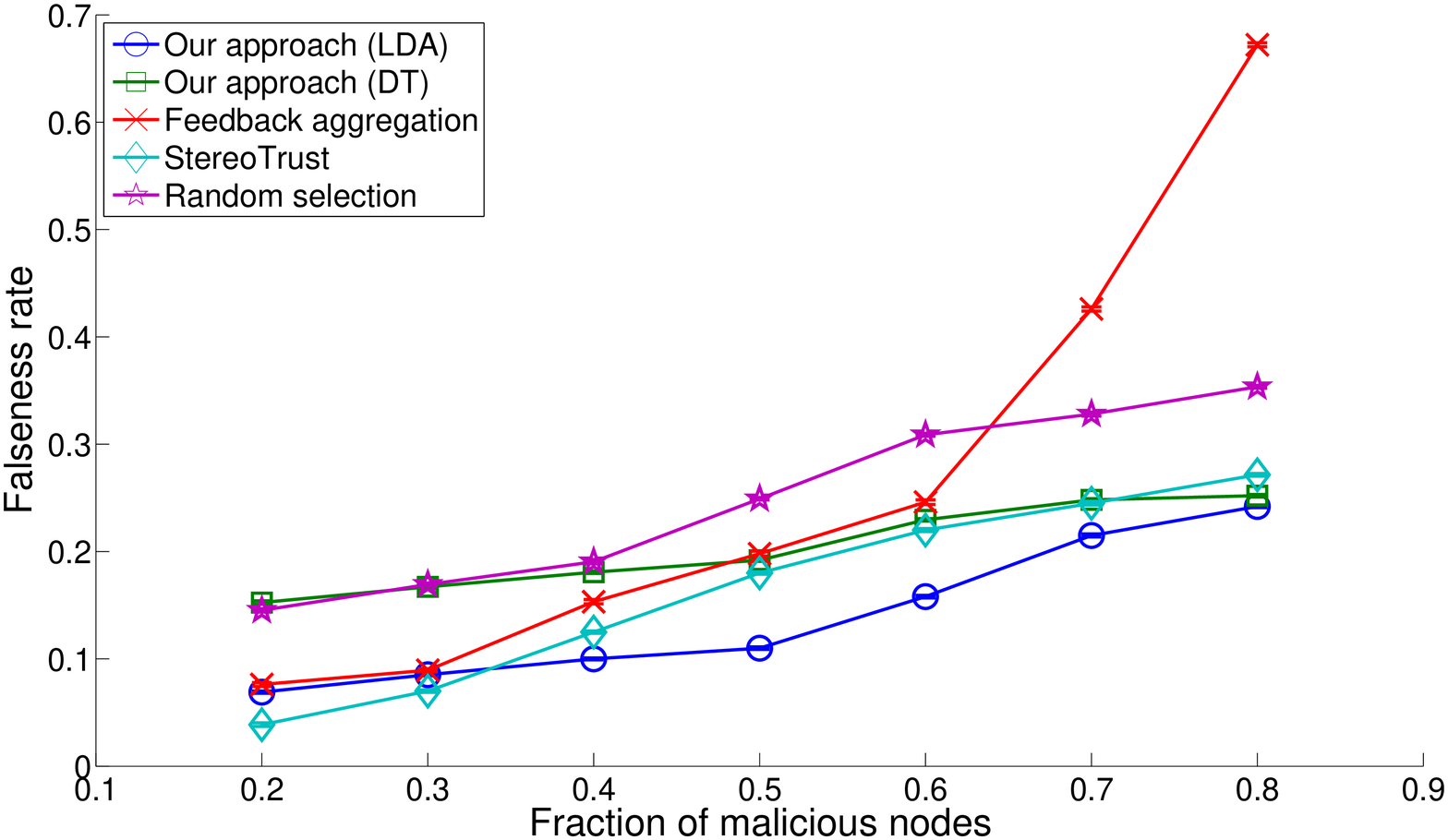}}\vspace{-0.1cm}\hspace*{-0.7cm}
    \subfigure[\label{fig:FN02}False negative.]{\includegraphics[scale=0.18]{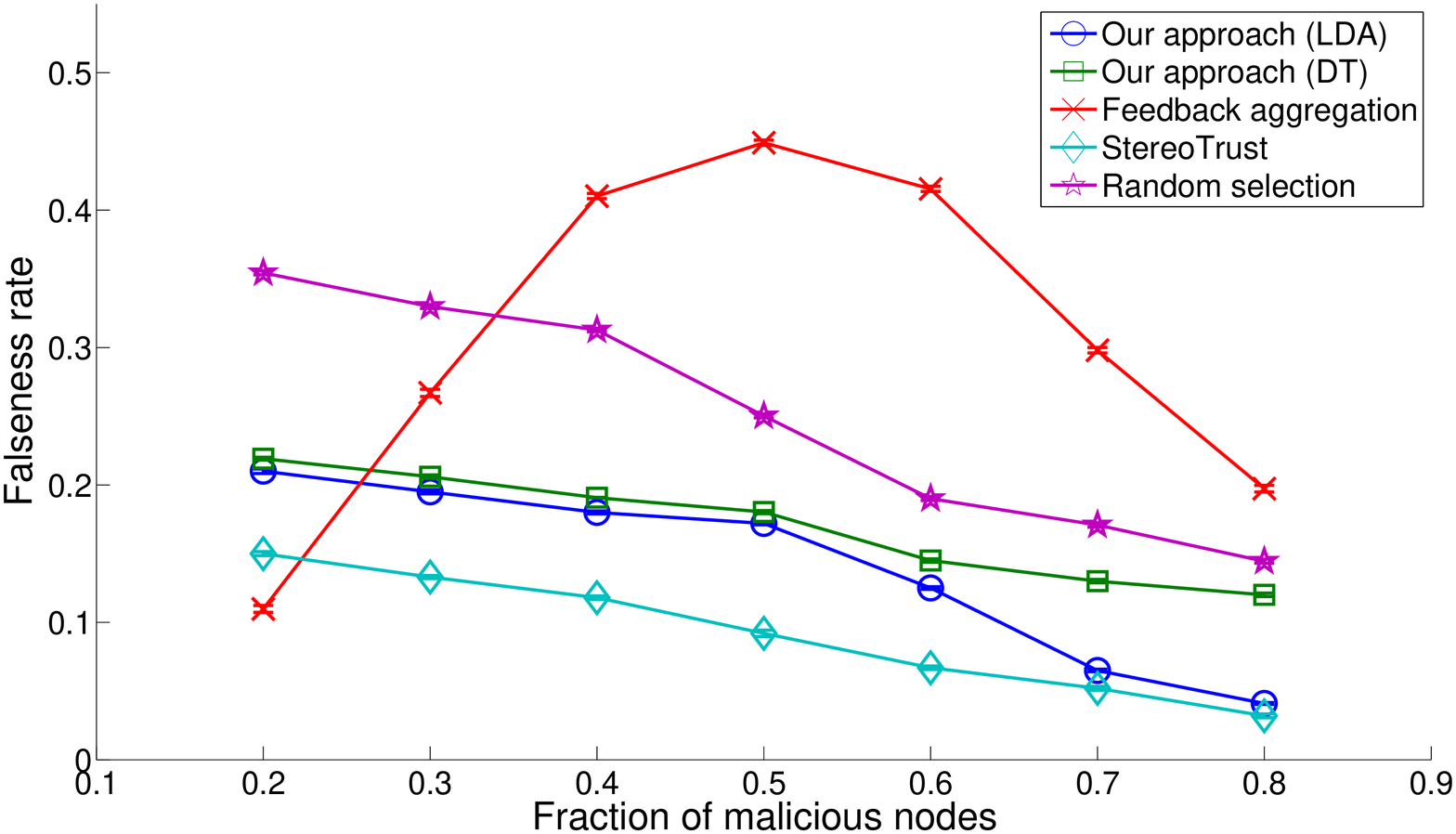}}\vspace{-0.1cm}\hspace*{-0.7cm}
    \subfigure[\label{fig:F02}Overall falseness.]{\includegraphics[scale=0.18]{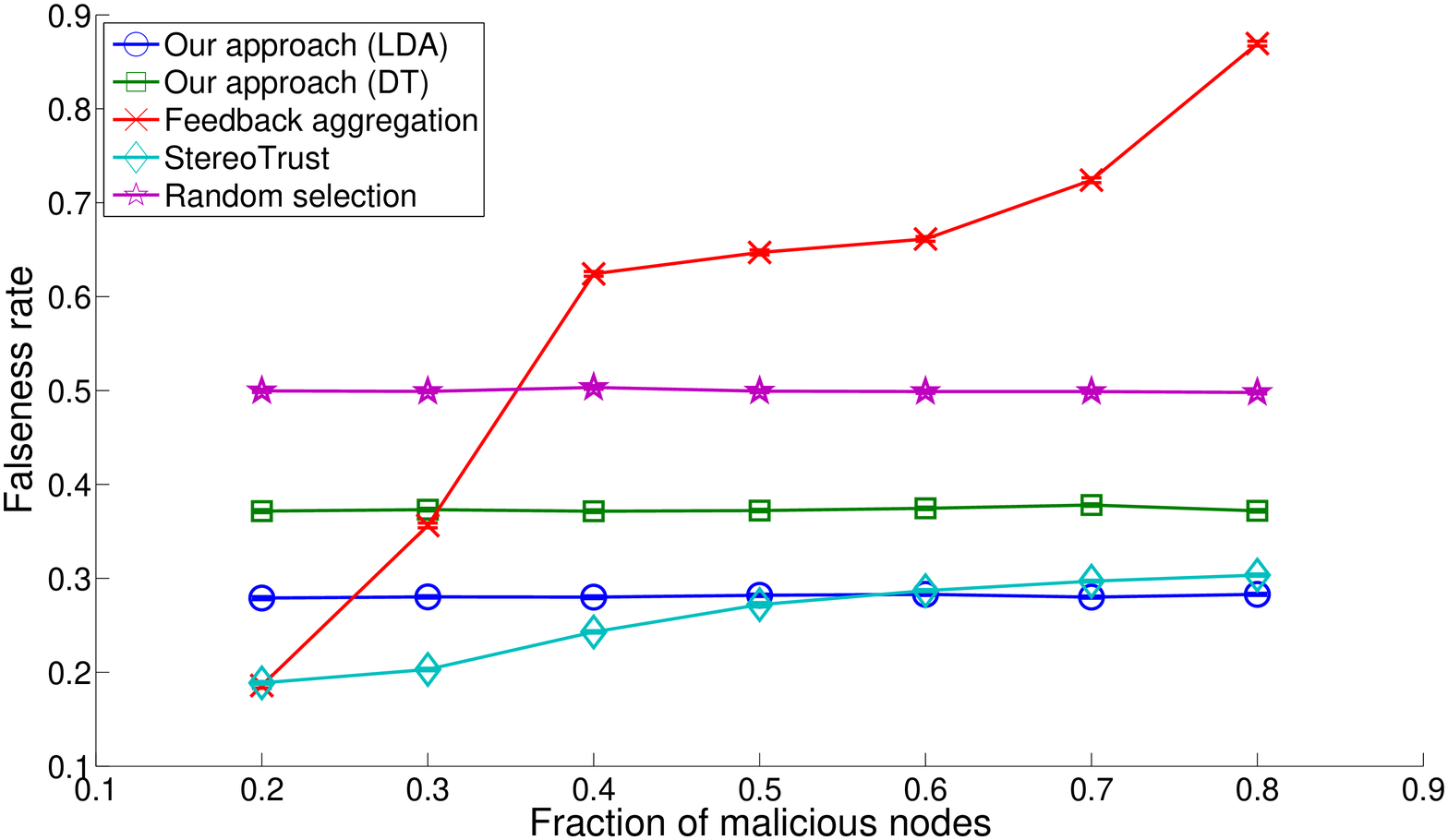}}\vspace{-0.1cm}
  \end{center}\vspace{-5mm}
  \caption{Comparison of our approach with other approaches ($\theta = 0.2$).}\vspace{-0.4cm}
\label{fig:compare02}
\end{figure}

\begin{figure}[tbp]
  \begin{center}
  \hspace*{-3.2cm}
  \centering
    \subfigure[\label{fig:FP04}False positive.]{\includegraphics[scale=0.18]{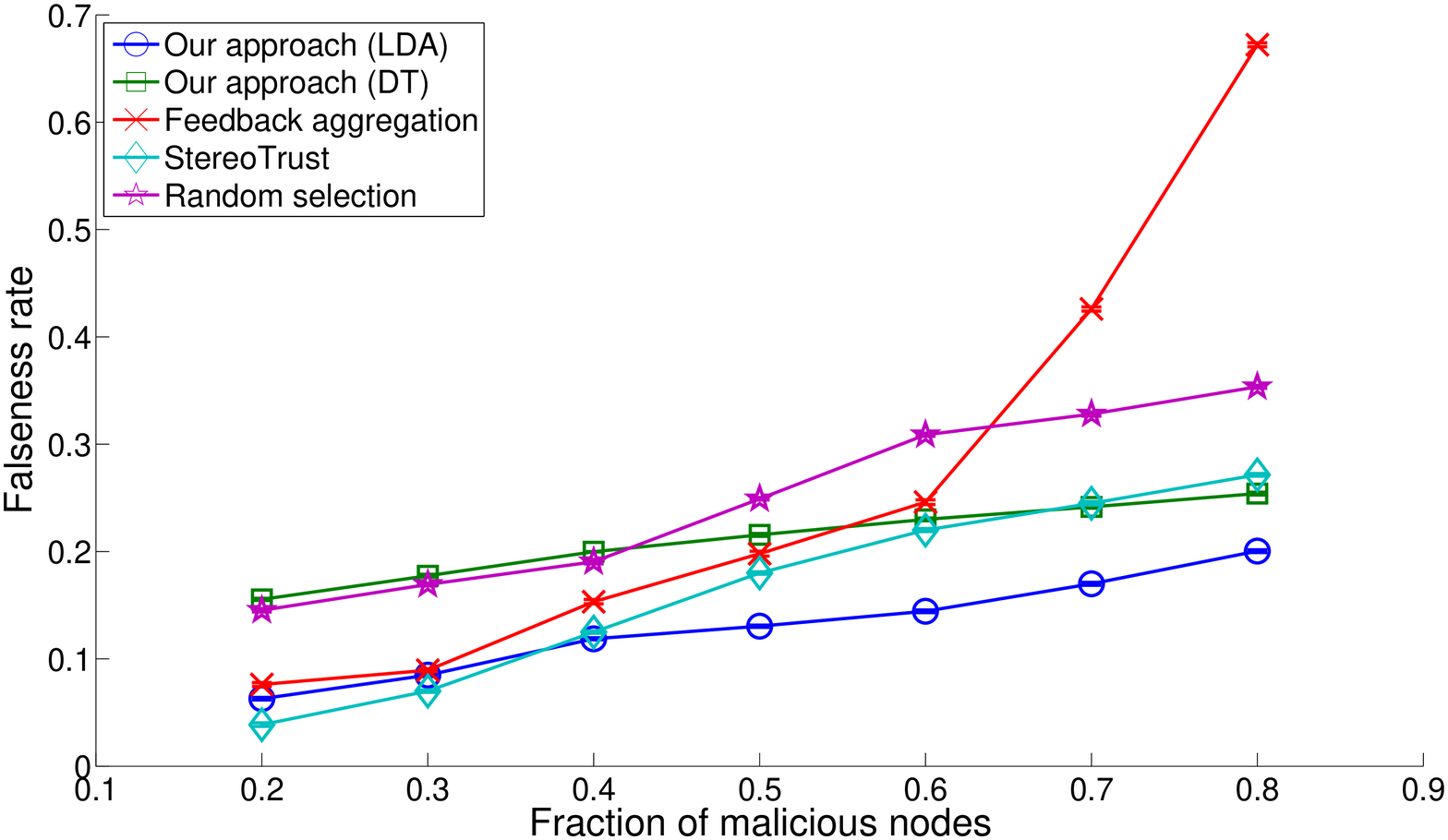}}\vspace{-0.1cm}\hspace*{-0.7cm}
    \subfigure[\label{fig:FN04}False negative.]{\includegraphics[scale=0.18]{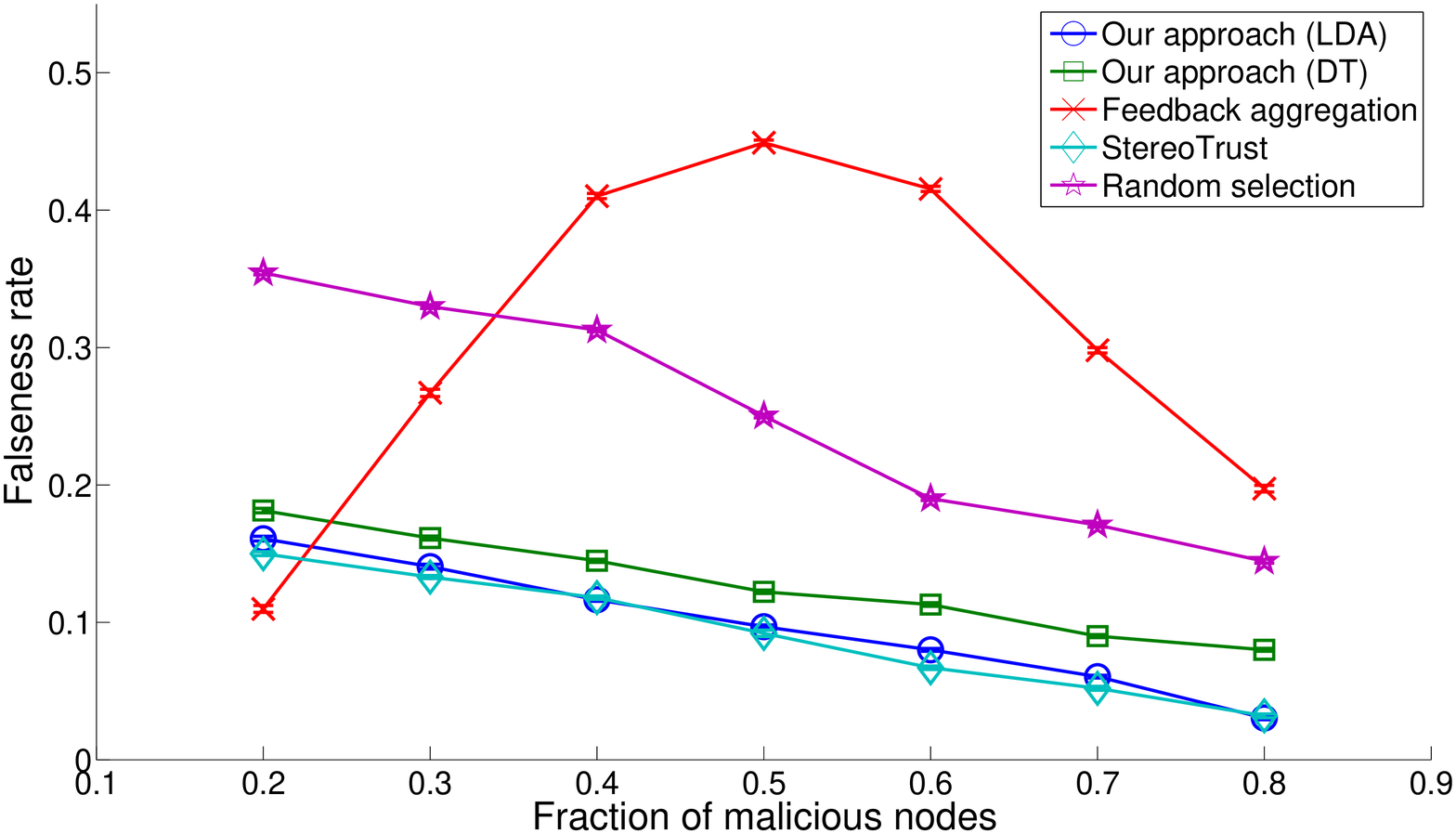}}\vspace{-0.1cm}\hspace*{-0.7cm}
    \subfigure[\label{fig:F04}Overall falseness.]{\includegraphics[scale=0.18]{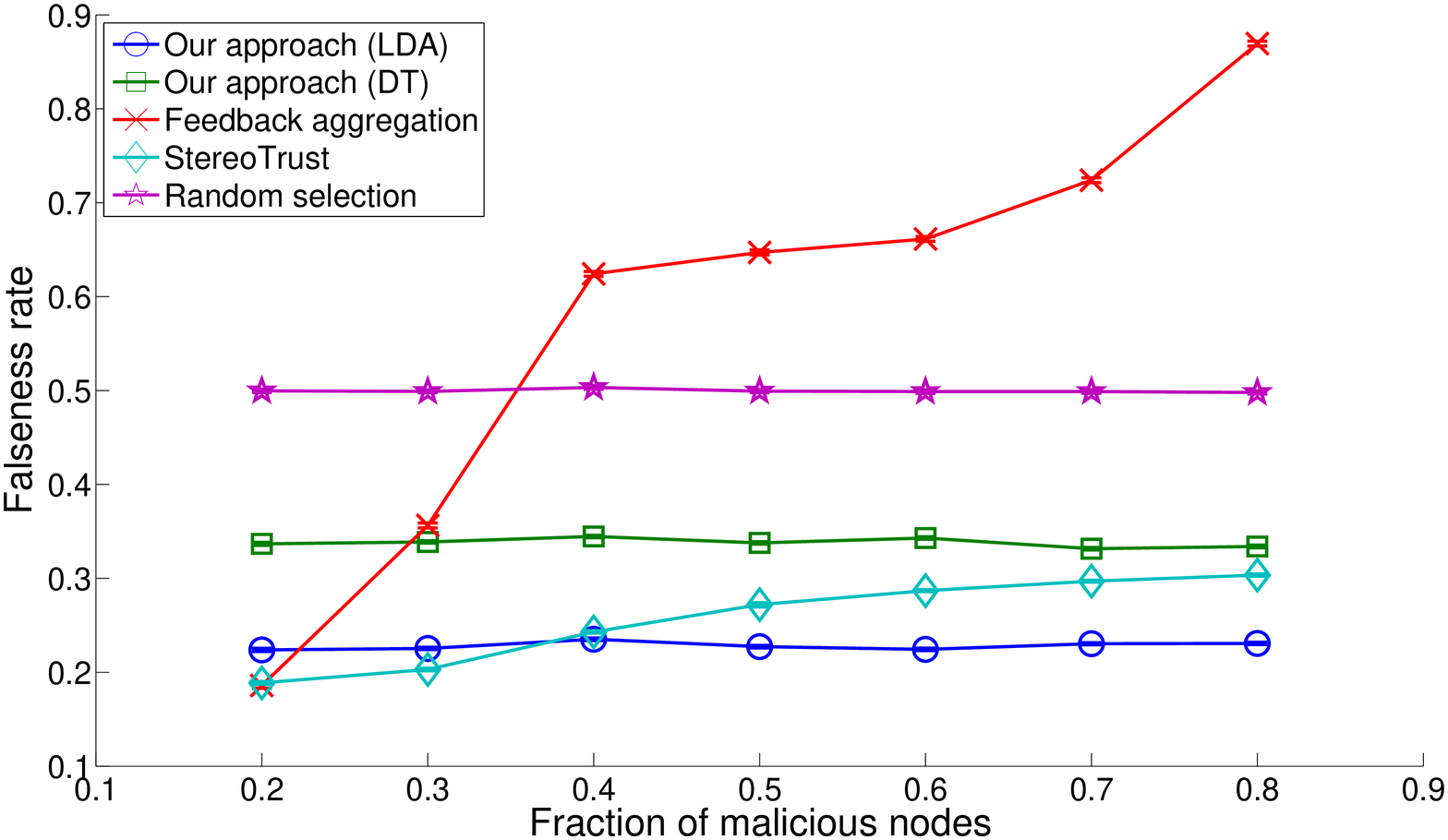}}\vspace{-0.1cm}
  \end{center}\vspace{-5mm}
  \caption{Comparison of our approach with other approaches ($\theta = 0.4$).}\vspace{-0.4cm}
  \vspace{-1mm}
\label{fig:compare04}
\end{figure}

\begin{figure}[tbp]
  \begin{center}
  \hspace*{-3.2cm}
  \centering
    \subfigure[\label{fig:FP06}False positive.]{\includegraphics[scale=0.18]{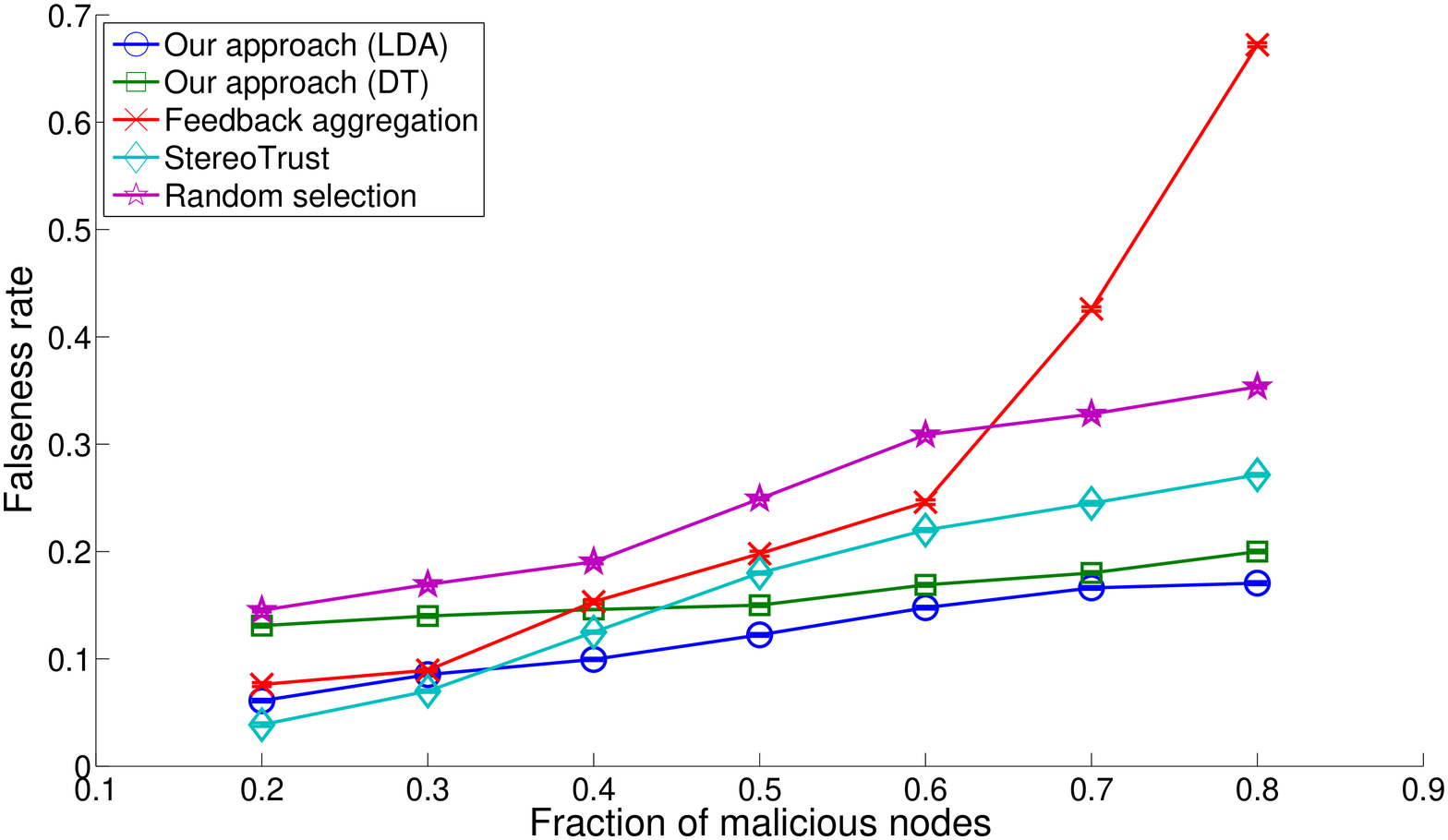}}\vspace{-0.1cm}\hspace*{-0.7cm}
    \subfigure[\label{fig:FN06}False negative.]{\includegraphics[scale=0.18]{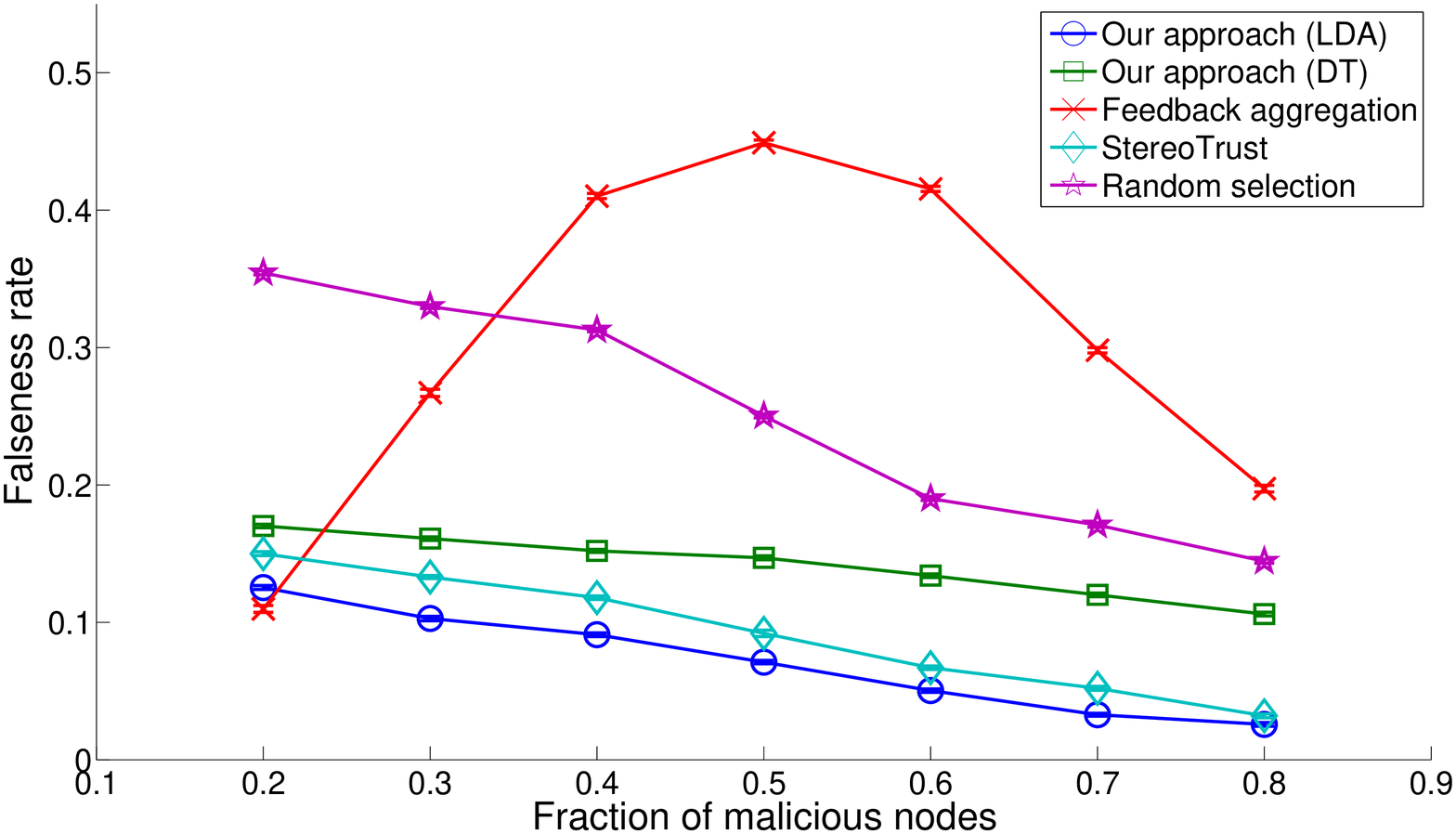}}\vspace{-0.1cm}\hspace*{-0.7cm}
    \subfigure[\label{fig:F06}Overall falseness.]{\includegraphics[scale=0.18]{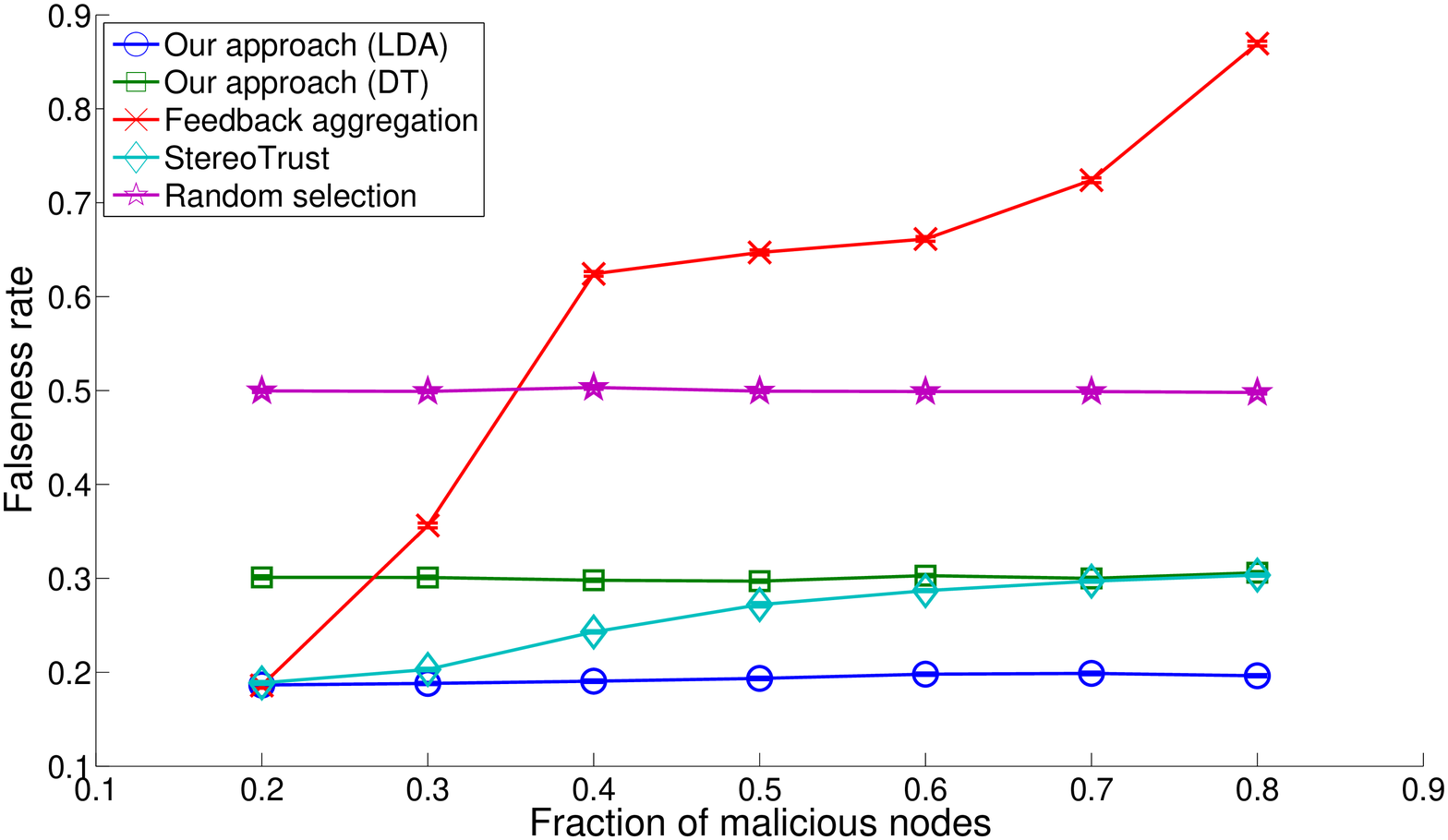}}\vspace{-0.1cm}
  \end{center}\vspace{-5mm}
  \caption{Comparison of our approach with other approaches ($\theta = 0.6$).}\vspace{-0.4cm}
\label{fig:compare06}
\end{figure}

\begin{figure}[tbp]
  \begin{center}
  \hspace*{-3.2cm}
  \centering
    \subfigure[\label{fig:FP08}False positive.]{\includegraphics[scale=0.18]{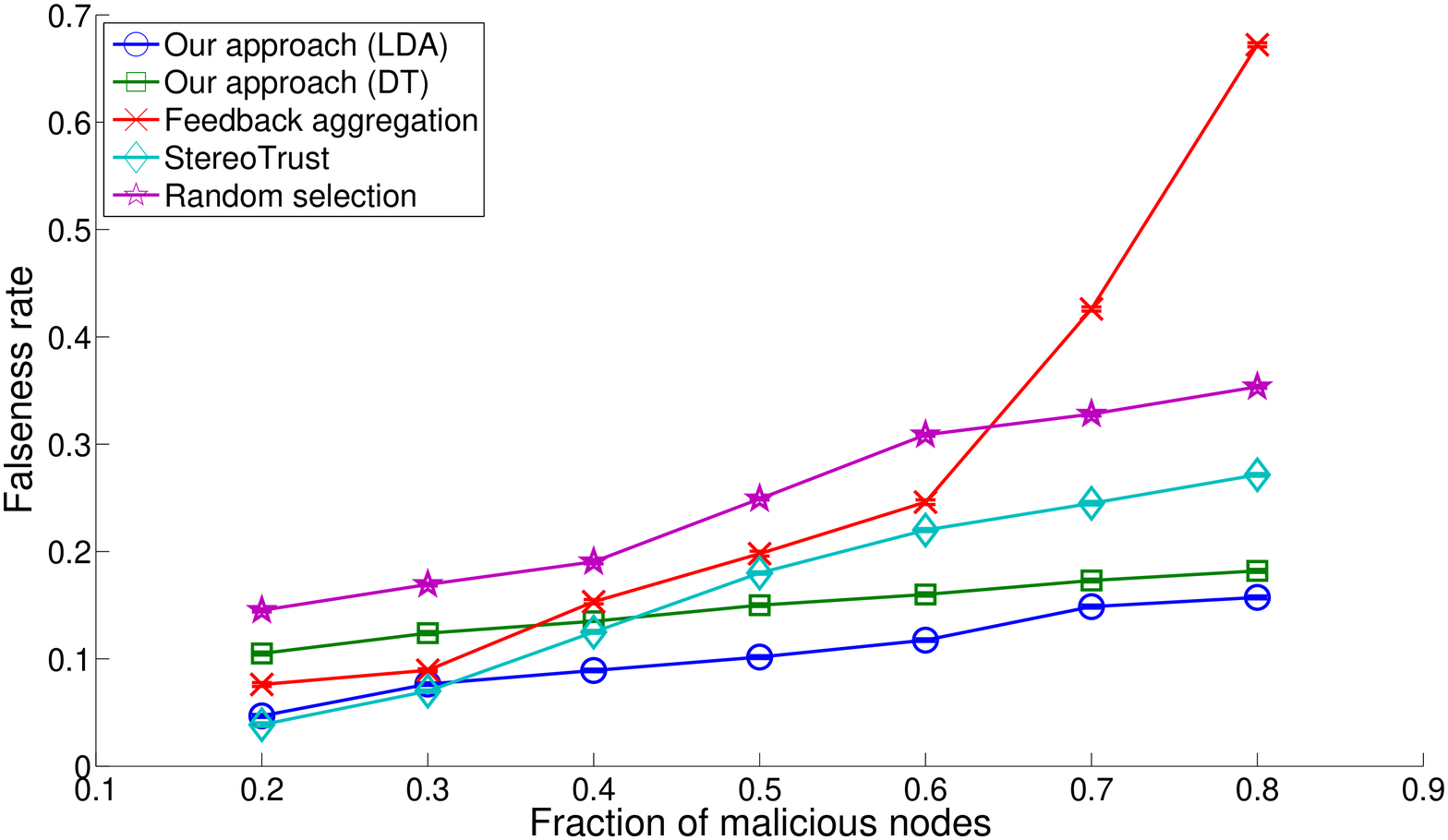}}\vspace{-0.1cm}\hspace*{-0.7cm}
    \subfigure[\label{fig:FN08}False negative.]{\includegraphics[scale=0.18]{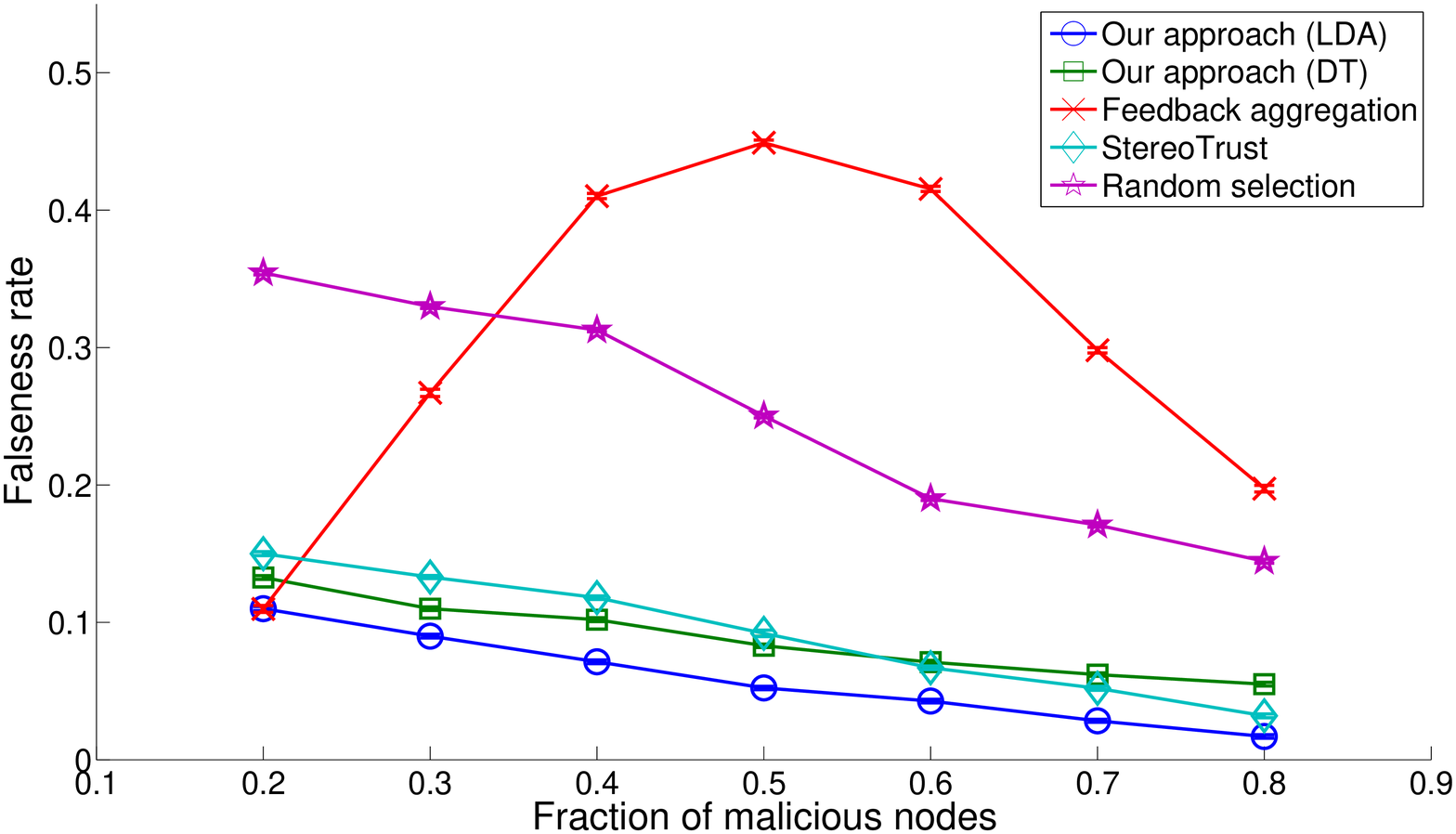}}\vspace{-0.1cm}\hspace*{-0.7cm}
    \subfigure[\label{fig:F08}Overall falseness.]{\includegraphics[scale=0.18]{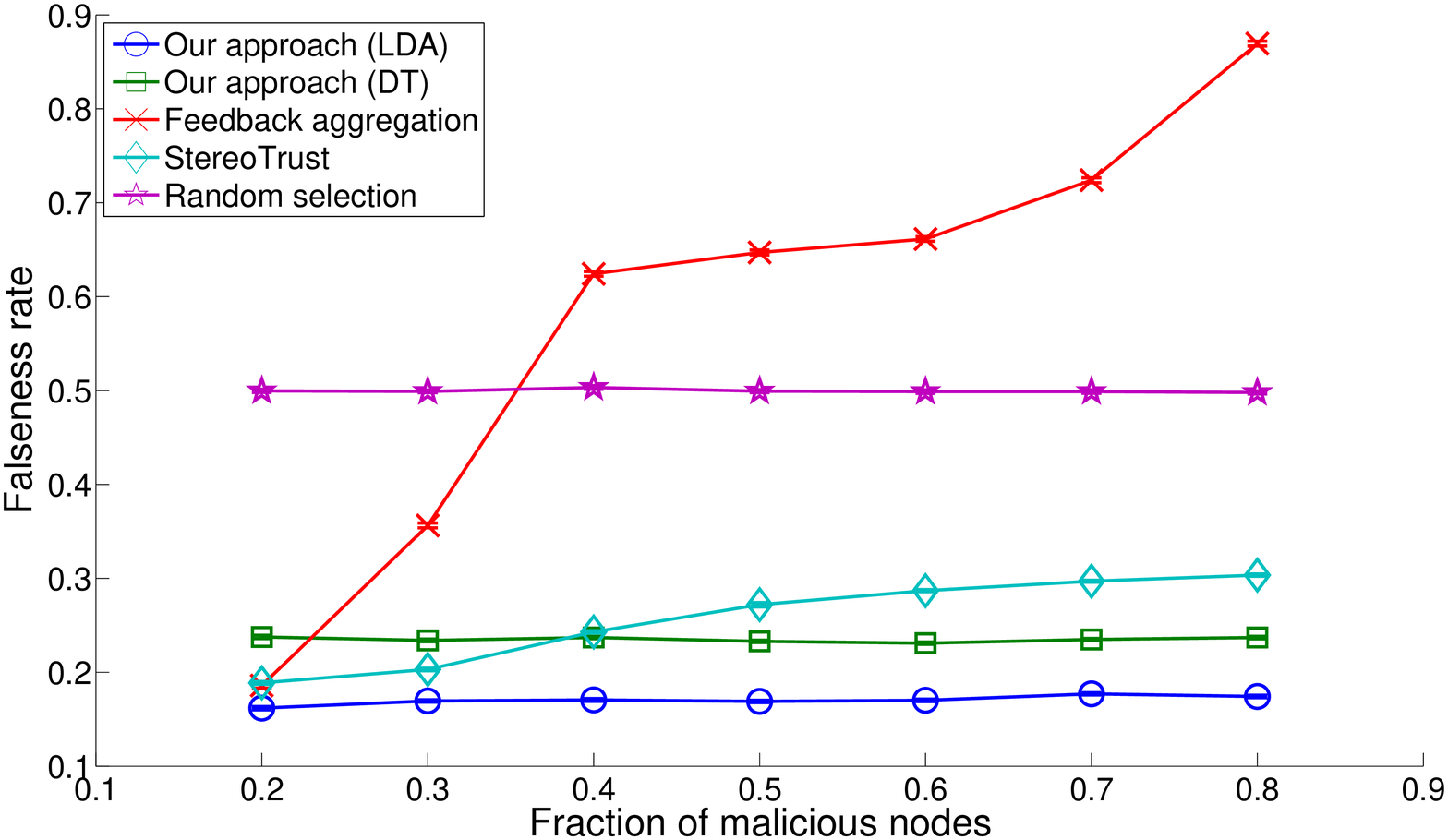}}\vspace{-0.1cm}
  \end{center}\vspace{-5mm}
  \caption{Comparison of our approach with other approaches ($\theta = 0.8$).}\vspace{-0.4cm}
\label{fig:compare08}
\end{figure}

We compare performance of our approach with other existing trust models
(Fig. \ref{fig:compare02}, \ref{fig:compare04}, \ref{fig:compare06} and \ref{fig:compare08}).
Notice that when local knowledge is sufficient (i.e., at least three successful transactions and three unsuccessful transactions), trustor relies on its local knowledge to perform our approach. Otherwise, trustor will integrate third party information via LKSON.

We observe that when
$P_{m}$ is low, feedback aggregation model performs quite well. This is because
when most of nodes in the system are honest, the aggregated feedbacks can
accurately predict behavior of the service provider thus keeping rates of false
positive, false negative as well as overall falseness low. However, with the
increase of $P_{m}$, various falseness increases sharply. So accuracy of
feedback aggregation model heavily depends on percentage of malicious nodes and
thus is not suitable for a hostile environment.

One interesting thing about
feedback aggregation model is that its false negative rate decreases when
$P_{m}$ is larger than 0.5 (Fig. \ref{fig:FN02}, \ref{fig:FN04}, \ref{fig:FN06} and
\ref{fig:FN08}). The possible reason is that when $P_{m}$ is high, the
amount of false feedbacks increases thus increasing the falseness
probability. On the other hand, the total amount of successful transactions
decreases generally, so the false negative rate decreases.

Random selection model, as we expected, provides a stable rate of overall
falseness of 0.5 because we let trustor accept or reject transactions equiprobably. This model is the benchmark of the experiments, demonstrating falseness rate evolution in the scenario where no trust management is applied.

StereoTrust model outperforms feedback aggregation in general especially when $P_{m}$ is high. This is because even though StereoTrust also relies on some feedbacks to predict behavior of the target node, it only requests nodes that are honest (from its own perspective) thus the probability of receiving false feedbacks is relatively low.

We observe that DT based approach is less accurate than LDA based approach in all scenarios. This result conforms to the result when real data is used (Fig. \ref{fig:varyingbuyersreal}) and further proves that LDA based algorithm is more suitable to the scenarios where features are quantitative.
So to sum up, LDA based approach, which has the lowest rate of various falseness outperforms all other models (except for the scenarios that $\theta$ is quite small, i.e., when $\theta \leq 0.4$). When $\theta$ is quite small, both LDA based approach and DT based approach suffer from low accuracy of classification.
Finding features that can distinguish successful transactions from unsuccessful ones is thus the key of our approach.  We also notice that performance of our approaches are quite stable (not affected by $P_{m}$ very much). This is because our approach mainly relies on local knowledge, which is the most reliable information source. Note that since our approach relies on different kinds of
information, it may not be rather fair to compare with other models that only use one kind of information (i.e., target agent's historical information). Moreover, in some application scenarios, the information used by our approach (e.g., features) may not be always available. Nevertheless, whenever such information is there, our approach indeed produces reasonable results thus demonstrating its efficiency.

\section{Related Work}
\label{sec:relatedwork}

Most existing reputation based trust management mechanisms derive an agent's trustworthiness
based on that specific agent's past behavior (e.g., \cite{distributedtrust97, Mui02acomputational, eigen03, p2prep, gossiptrust, probabilistictrustreputation, hybridtransitivetrust}). If the assessing agent
has sufficient direct experience with a provider agent, this provider's future
behavior can be reliably predicted \cite{Mui02acomputational}. However, in
large-scale distributed settings, direct experience is often unavailable. In
this case, the assessing agent may resort to ``indirect experience'' -- opinions
about the assessed agent obtained from other agents \cite{distributedtrust97, betareputation02, eigen03, Xiong04peertrust, probabilistictrustreputation}. Simple aggregation (like a
seller's ranking on eBay) relies on access to global information like history of
the assessed agent's behavior. Alternatively, transitive trust models
\cite{distributedtrust97, transitive03} forms chains of trust relationships
among agents.
However, transitive trust is not always realistic in the real deployment and it has several drawbacks: (i) this method does not handle wrong recommendations properly, which affect the accuracy of derived trust seriously \cite{Datta03beyondweb}. (ii) This method does not provide a
mechanism for updating trust efficiently in a dynamic system. (iii) Establishing
a trust path, even if such a path exists, is nontrivial.

EigenTrust \cite{eigen03} is a reputation system developed for P2P networks.
EigenTrust uses transitivity of trust and aggregates trust from peers by
performing a distributed calculation to determine the eigenvector of a ``trust
matrix'' over peers. It relies on some pre-trusted peers, which are supposed to
be trusted by all peers. EigenTrust (and some other reputation systems like
\cite{Xiong04peertrust,powertrust}) is designed based on Distributed Hash Tables
(DHT) \cite{chord, pastry, pgrid} thus imposing system design complexity and drawbacks, deployment and message overheads \cite{nontransitivedht}. \cite{eigengroup} proposed a trust
system that is built on top of a peer group infrastructure using an
EigenTrust-like calculation approach. The groups are formed based on particular
interest criterions and members must follow the set of rules of the group they
belong to. The authors assumed that a group leader creates the group and
controls the membership. To calculate trust, the authors introduced Eigen Group
Trust, which is an aggregative version of EigenTrust. In Eigen
Group trust, all the transactions rely on the group leaders, who are assumed to
be trusted and resourceful.

GossipTrust \cite{gossiptrust} is another reputation management system designed for P2P networks. More specifically, it is designed for unstructured P2P networks. The basic idea of GossipTrust is to aggregate global reputation scores using a gossip based algorithm. The gossiping process stops when a gossiping error threshold is reached. Such approach has the advantage of a fast dissemination of global scores with time complexity of $O(log_2(n))$, where $n$ is the number of peers in the network.

The Beta Reputation System (BRS) \cite{betareputation02} is a probabilistic
trust model, which is based on the beta distribution. In BRS, after each interaction, service requester gives rating to performance of the service provider, where rating is a binary
value: positive or negative. The ratings are then used to estimate shape
parameters that determine the reputation of the assessed agent.
However, BRS does not show how it is able to cope with false information, which
may influence accuracy of prediction seriously.

TRAVOS \cite{travos06} is a trust and reputation model for agent-based virtual organizations. This work also makes use of beta distribution to compute trust but it pays more attention to the issue of unfair feedbacks. When a buyer evaluates a potential interaction partner (i.e., seller), it first uses its own local knowledge to derive trustworthiness of the seller. It then estimates accuracy of the direct trust, that is the probability that the real likelihood of cheating falls in the certain range from the buyer's estimation. If the direct trust is accurate enough, the buyer simply relies on this direct experience based trust. Otherwise, it requests feedbacks from other buyers who have interacted with the potential seller. To ensure that only accurate feedbacks are considered, TRAVOS addresses inaccurate reputation feedbacks by performing two tasks:
(1) Estimating the probability that a feedback provider's opinion of the seller is accurate. This is done by comparing current feedback with the previous feedbacks provided by the same buyer. The accuracy of the current feedback is the expected value of the beta PDF representing the numbers of successful and unsuccessful interactions between the buyer and the seller when the buyer is guided by the previous feedbacks.
(2) Adjusting reputation feedback according to its accuracy to reduce the effect of inaccurate feedback.

Different from works mentioned above, StereoTrust \cite{stereotrust09} uses
another kind of information, i.e., stereotypes to estimate the initial trust of an
unknown agent. A stereotype is determined by a feature vector, which can be taken
from the profile of the agent. To build stereotypes, the agents that
trustor has interacted with are classified into groups according to the
identified features. Stereotypes on each group are calculated by aggregating trustor's past experience with members of that group. Then, when facing the target agent, the trustor estimates its trust using stereotypes on groups to which the target agent belongs. Additionally, when some information about stranger's past behavior is available, enhanced StereoTrust model (called d-StereoTrust) uses it to refine the stereotype matching.
StereoTrust simply aggregates stereotypes by assigning intuitive weights to derive trust of the target agent, thus is not able to tell which stereotypes are more important than other ones. Differently, by applying sophisticated machine learning algorithms, our approach is capable of distinguishing which feature is more discriminating thus making the prediction more accurate.

Similar to StereoTrust, \cite{bootstrapStereotype} proposed to bootstrap trust of an unknown agent through stereotypes, which are estimated based on M5 model tree learning algorithm \cite{Quinlan92learningwith}. This work also discussed how to combine initial trust (i.e., based on stereotypes) and reputation based trust using subjective logic \cite{subjectivelogic}. Differently, our approach focuses on transaction instead of individual agent. This helps address the inconsistency of the agent behavior since its target is not individual agents but overall viewpoint on a group of
agents that share the similar properties. Moveover, our approach emphasizes the useful feature selection by performing machine learning algorithms and demonstrates its efficacy in a realistic setting using a real auction dataset.

\section{Conclusion}
\label{sec:conclusion}
In this paper, we propose a generic trust framework, which is designed for large-scale, open systems. Our approach is generically based on machine learning algorithms, which provide classification of a potential transaction by learning a set of past transactions. In this work we instantiate the framework using two common machine learning algorithms, i.e., linear discriminant analysis and decision tree. Unlike many existing trust mechanisms, which rely on specific agent's historical information to predict its future behavior, our approach only uses trustor's local knowledge. This makes our approach quite suitable for a large-scale distributed environment where historical information of the agent in question is scarce or even not available, and third party information about the agent may not be reliable or expensive to obtain.

In case that trustor does not have sufficient local knowledge, we propose to construct a local knowledge sharing overlay network (LKSON) to exchange agents' local knowledge (i.e., intermediate results of classification algorithms) to predict trustworthiness of a potential transaction. Compared to traditional feedback aggregation based trust and reputation systems, such mechanism helps (1) reduce the possibility of sharing fake information; (2) avoid privacy leakage, (3) save a lot of computation and (4) avoid high communication overhead.

Not surprisingly, simulation results show that performance of our approach is
positively correlated with the discrimination power of features on
successful interactions and unsuccessful interactions.
However, compared to other trust models, our approach is quite efficient,
especially when third party information is not reliable. Moreover, performance
of our approach is quite stable because it only relies on trustor's local
knowledge. The real Allegro dataset based simulation shows that the proposed approach can be applied to real application (e.g., detecting Internet auction frauds).

In the future works, we are going to investigate more sophistical statistical tools (e.g., Multiple Discriminant Analysis, etc.) to further improve accuracy of our approach. Another direction is to apply the proposal to some concrete applications such as recommending interesting contents in web 2.0 sites (e.g., YouTube\footnote{http://www.youtube.com}, Digg\footnote{http://digg.com/}, etc.).

\section{Acknowledgement}
\label{sec:ack}
The authors are grateful to Dr. Adam Wierzbicki of Polish Japanese Institute of Information Technology for his assistance in providing Allegro dataset.


\begin{thebibliography}{10}

\bibitem{distributedtrust97}
Alfarez Abdul-Rahman and Stephen Hailes.
\newblock A distributed trust model.
\newblock In {\em NSPW '97: Proceedings of the 1997 workshop on New security
  paradigms}, 1997.

\bibitem{pgrid}
Karl Aberer, Philippe Cudr\'{e}-Mauroux, Anwitaman Datta, Zoran Despotovic,
  Manfred Hauswirth, Magdalena Punceva, and Roman Schmidt.
\newblock P-grid: a self-organizing structured p2p system.
\newblock {\em SIGMOD Rec.}, 32(3):29--33, 2003.

\bibitem{allegro}
Allegro.
\newblock http://allegro.pl/, 2011.

\bibitem{p2prep}
R.~Aringhieri, E.~Damiani, S.~De, Capitani~Di Vimercati, S.~Paraboschi, and
  P.~Samarati.
\newblock Fuzzy techniques for trust and reputation management in anonymous
  peer-to-peer systems.
\newblock {\em Journal of the American Society for Information Science and
  Technology}, 57:528--537, 2006.

\bibitem{bootstrapStereotype}
Chris Burnett, Timothy~J. Norman, and Katia Sycara.
\newblock Bootstrapping trust evaluations through stereotypes.
\newblock In {\em Proceedings of 9th International Joint Conference on
  Autonomous Agents and Multiagent Systems (AAMAS)}, 2010.

\bibitem{Datta03beyondweb}
Anwitaman Datta, Manfred Hauswirth, and Karl Aberer.
\newblock Beyond "web of trust": Enabling p2p e-commerce.
\newblock In {\em Proceedings of the IEEE CEC}, 2003.

\bibitem{epinions}
Epinions.com.
\newblock http://www.epinions.com/, 2011.

\bibitem{nontransitivedht}
Michael~J. Freedman, Karthik Lakshminarayanan, Sean Rhea, and Ion Stoica.
\newblock Non-transitive connectivity and dhts.
\newblock In {\em WORLDS'05: Proceedings of the 2nd conference on Real, Large
  Distributed Systems}, 2005.

\bibitem{fukunaga1990}
Keinosuke Fukunaga.
\newblock {\em Introduction to Statistical Pattern Recognition}.
\newblock Academic Press Professional, Inc., 1990.

\bibitem{canwetrusttrust}
Diego Gambetta.
\newblock Can we trust trust?
\newblock In D.~Gambetta, editor, {\em Trust: Making and Breaking Cooperative
  Relations}. Basil Blackwell, 1988.

\bibitem{dtNPhard}
Laurent Hyafil and R.~L. Rivest.
\newblock {Constructing Optimal Binary Decision Trees is NP-complete}.
\newblock {\em Information Processing Letters}, 5(1):15--17, 1976.

\bibitem{subjectivelogic}
Audun J{\o}ang.
\newblock Artificial reasoning with subjective logic.
\newblock In {\em Proceedings of the Second Australian Workshop on Commonsense
  Reasoning}, 1997.

\bibitem{transitive03}
A.~J{\o}sang, E.~Gray, and M.~Kinateder.
\newblock Analysing topologies of transitive trust.
\newblock In {\em Proceedings of the Workshop of Formal Aspects of Security and
  Trust (FAST)}, 2003.

\bibitem{betareputation02}
Audun J{\o}sang and Roslan Ismail.
\newblock The beta reputation system.
\newblock In {\em Proceedings of the 15th Bled Conference on Electronic
  Commerce}, 2002.

\bibitem{josongsurvey}
Audun J{\o}sang, Roslan Ismail, and Colin Boyd.
\newblock A survey of trust and reputation systems for online service
  provision.
\newblock {\em Decis. Support Syst.}, 43:618--644, 2007.

\bibitem{eigen03}
Sepandar~D. Kamvar, Mario~T. Schlosser, and Hector Garcia-Molina.
\newblock The eigentrust algorithm for reputation management in p2p networks.
\newblock In {\em WWW '03: Proceedings of the 12th international conference on
  World Wide Web}, 2003.

\bibitem{stereotrust09}
Xin Liu, Anwitaman Datta, Krzysztof Rzadca, and Ee-Peng Lim.
\newblock Stereotrust: a group based personalized trust model.
\newblock In {\em CIKM '09: Proceeding of the 18th ACM conference on
  Information and knowledge management}, 2009.

\bibitem{informationtheory}
David J.~C. MacKay.
\newblock {\em Information theory, inference, and learning algorithms}.
\newblock Cambridge University Press, 2003.

\bibitem{da}
Geoffrey~J. McLachlan.
\newblock {\em Discriminant Analysis and Statistical Pattern Recognition}.
\newblock Wiley-Interscience, Augest 2004.

\bibitem{mlintroduction}
Tom Mitchell.
\newblock {\em Machine Learning}.
\newblock McGraw Hill, 1997.

\bibitem{Mui02acomputational}
Lik Mui and Mojdeh Mohtashemi.
\newblock A computational model of trust and reputation.
\newblock In {\em Proceedings of the 35th HICSS}, 2002.

\bibitem{introDT}
J.~R. Quinlan.
\newblock Induction of decision trees.
\newblock {\em Mach. Learn.}, 1(1):81--106, 1986.

\bibitem{Quinlan92learningwith}
J.~R. Quinlan.
\newblock Learning with continuous classes.
\newblock In {\em Proceedings of the 5th Australian Joint Conference on
  Artificial Intelligence}, pages 343--348, 1992.

\bibitem{c45}
J.~Ross Quinlan.
\newblock {\em C4.5: programs for machine learning}.
\newblock Morgan Kaufmann Publishers Inc., San Francisco, CA, USA, 1993.

\bibitem{eigengroup}
Ajay Ravichandran and Jongpil Yoon.
\newblock Trust management with delegation in grouped peer-to-peer communities.
\newblock In {\em SACMAT '06: Proceedings of the eleventh ACM symposium on
  Access control models and technologies}, pages 71--80, New York, NY, USA,
  2006. ACM.

\bibitem{pastry}
Antony I.~T. Rowstron and Peter Druschel.
\newblock Pastry: Scalable, decentralized object location, and routing for
  large-scale peer-to-peer systems.
\newblock In {\em Middleware '01: Proceedings of the IFIP/ACM International
  Conference on Distributed Systems Platforms Heidelberg}, 2001.

\bibitem{chord}
Ion Stoica, Robert Morris, David Karger, M.~Frans Kaashoek, and Hari
  Balakrishnan.
\newblock Chord: A scalable peer-to-peer lookup service for internet
  applications.
\newblock In {\em SIGCOMM '01: Proceedings of the 2001 conference on
  Applications, technologies, architectures, and protocols for computer
  communications}, 2001.

\bibitem{Swamynathan05decouplingservice}
Gayatri Swamynathan, Ben~Y. Zhao, and Kevin~C. Almeroth.
\newblock Decoupling service and feedback trust in a peer-to-peer reputation
  system.
\newblock In {\em In Proc. of International Workshop on Applications and
  Economics of Peer-to-Peer Systems (AEPP)}, pages 82--90. Springer-Verlag,
  2005.

\bibitem{hybridtransitivetrust}
Jie Tang, Sven Seuken, and David~C. Parkes.
\newblock Hybrid transitive trust mechanisms.
\newblock In {\em Proceedings of 9th International Joint Conference on
  Autonomous Agents and Multiagent Systems (AAMAS)}, 2010.

\bibitem{travos06}
W.~T.~L. Teacy, J.~Patel, N.~R. Jennings, and M.~Luck.
\newblock Travos: Trust and reputation in the context of inaccurate information
  sources.
\newblock {\em Autonomous Agents and Multi-Agent Systems}, 12:183--198, 2006.

\bibitem{probabilistictrustreputation}
George Vogiatzis, Ian MacGillivray, and Maria Chli.
\newblock A probabilistic model for trust and reputation.
\newblock In {\em Proceedings of 9th International Joint Conference on
  Autonomous Agents and Multiagent Systems (AAMAS)}, 2010.

\bibitem{Xiong04peertrust}
Li~Xiong and Ling Liu.
\newblock Peertrust: Supporting reputation-based trust for peer-to-peer
  electronic communities.
\newblock {\em IEEE Transactions on Knowledge and Data Engineering},
  16:843--857, 2004.

\bibitem{powertrust}
Runfang Zhou and Kai Hwang.
\newblock Powertrust: A robust and scalable reputation system for trusted
  peer-to-peer computing.
\newblock {\em IEEE Trans. Parallel Distrib. Syst.}, 18:460--473, 2007.

\bibitem{gossiptrust}
Runfang Zhou, Kai Hwang, and Min Cai.
\newblock Gossiptrust for fast reputation aggregation in peer-to-peer networks.
\newblock {\em IEEE Transactions on Knowledge and Data Engineering},
  20:1282--1295, 2008.

\end{thebibliography}

\end{document}